\documentclass[amsmath,preprintnumbers,10pt,twocolumn,prl]{revtex4-1}
\usepackage{amsbsy}
\usepackage{graphicx}
\usepackage{color}
\usepackage{subfigure}
\usepackage{physics}
\usepackage{soul}
\usepackage{color}
\usepackage{bm}

\usepackage{verbatim}
\usepackage{natbib}
\begin{document}
\title{Topological localization in out-of-equilibrium dissipative systems}
\author{Kinjal Dasbiswas$^1$, Kranthi K. Mandadapu$^{2,3}$, Suriyanarayanan Vaikuntanathan$^{1,4}$} 
\affiliation{$^1$The James Franck Institute, The University of Chicago, Chicago, IL,}
\affiliation{$^2$ Department of Chemical and Biomolecular Engineering, University of California, Berkeley, Berkeley, CA,}
\affiliation{$^3$ Chemical Sciences Division, Lawrence Berkeley National Laboratory, Berkeley, CA,}
\affiliation{$^4$ Department of Chemistry, The University of Chicago, Chicago, IL.}

\begin{abstract}

In this paper we report that notions of topological protection can be applied to stationary configurations that are driven far from equilibrium by active, dissipative processes. We show this for physically two disparate cases : stochastic networks governed by microscopic single particle dynamics as well as collections of driven, interacting particles described by coarse-grained hydrodynamic theory. In both cases, the presence of dissipative couplings to the environment that break time reversal symmetry are crucial to ensuring topologically protection.  These examples constitute proof of principle that notions of topological protection, established in the context of electronic and mechanical systems, do indeed extend generically to processes that operate out of equilibrium. Such topologically robust boundary modes have implications for both biological and synthetic systems.

\end{abstract}
\maketitle

Theoretical and experimental studies of biophysical mechanisms such as error correction in DNA replication~\cite{Hopfield1974}, adaptation in molecular motors controlling flagellar dynamics~\cite{Wang2017,Tu2008,Lan2012}, and timing of events in the cell cycle~\cite{Cao2015} are beginning to demonstrate the close connection between robust functioning and non-equilibrium forces. Theoretical results~\cite{Barato2015,Gingrich2016} have also elucidated the connection between energy dissipation and fluctuations in a wide class of non-equilibrium systems and have demonstrated how dissipation can be used to tune steady states of many body non-equilibrium systems~\cite{Nguyen2016}. However, unlike the behavior and characteristics of equilibrium systems, where no energy is dissipated, general principles governing fluctuations about a steady state or the steady state itself in far-from-equilibrium conditions are just being discovered. As such, strategies that allow for the engineering of specific robust steady states in non-equilibrium biological and soft matter systems are highly desirable. In this paper, we take an approach that is motivated by the physics of topological insulators to construct such states in non-equilibrium systems.

The discovery of robust localized edge states in mechanical systems \cite{Kane2013} that resemble those found in topologically non-trivial electronic \cite{Hasan2010} and photonic \cite{Haldane2008} systems has enabled the development of new design principles. For instance, it has been demonstrated that an assembly of coupled gyroscopes can support topologically protected directed modes at their boundary \cite{Nash2015}. Such assemblies can function as a robust waveguides. Localized edge modes have also been discovered in mechanical lattices \cite{Kane2013, Paulose2015}. These edge modes can be set up either at the edge of a lattice \cite{Kane2013} or at topological defects in the interior of the lattice \cite{Paulose2015}. They are formally zero energy ``free" modes but unlike the commonly found long wavelength zero energy modes in the bulk of marginally stable lattices, these topologically protected boundary modes are highly resistant to perturbations due to disorder and environmental fluctuations. Like their electronic and photonic counterparts, topological modes in the above mentioned mechanical systems have been characterized by topological indices such as winding or Chern numbers \cite{Ryu2002, Chen2014, Lubensky2015}.

%[Talk about how/why quantum mechanics protects fluctuations. Introduce concepts like winding number and band gaps.]
In this paper, we propose that topologically protected modes can be encoded in a variety of biological and soft matter contexts at the cost of energy dissipation.  Like in the electronic and mechanical analogs, the bulk of these systems can be characterized by a topological index (a winding number, in the cases we discuss), and the associated topologically protected zero modes we discover are localized at the edge or an interface. The modes are highly robust and insensitive to perturbations. In some specific contexts, the localization indirectly implies an edge current, analogous to that in the quantum Hall effect \cite{Hasan2010}. In all the instances considered in this paper, topological protection requires that the fundamental equations of motion contain dissipative couplings. 

We derive our results in two broad and apparently dissimilar contexts.  In the first, we consider biochemical networks with connectivity motivated by those of networks commonly encountered in biophysical information processing and control \cite{Murugan2014}. We show that the spectrum of the master equation rate matrix can support localized edge modes that are separated from the bulk via a band gap. In the second, we consider a hydrodynamic description of active matter, specifically that of collections of driven rotating particles in a confined geometry \cite{Tsai2005, vanZuiden16}, that exhibit large-scale flow localized at the boundaries.

The hydrodynamic equations have a structure very different from that of master equations considered in the first example.  The two address phenomena at very different length and time scales, one being a coarse-grained phenomenological description and the other a microscopic approach. Nonetheless, they both describe dissipative phenomena characterized by the production of entropy and lack time reversal symmetry. In fact, we find that they can be characterized by a topological index in the same way as introduced in the context of topological mechanics by Kane and Lubensky \cite{Kane2013}, that is by mapping to the Su-Schrieffer-Heeger (SSH) model for electrons on a 1D crystal lattice with two sites per unit cell \cite{SSH79} . Our results then elucidate the design principles required for robust steady states in various biophysical and synthetic systems.

\section{Topological protection in Markov networks}

\begin{figure}[h]
	\includegraphics[width=3.5in]{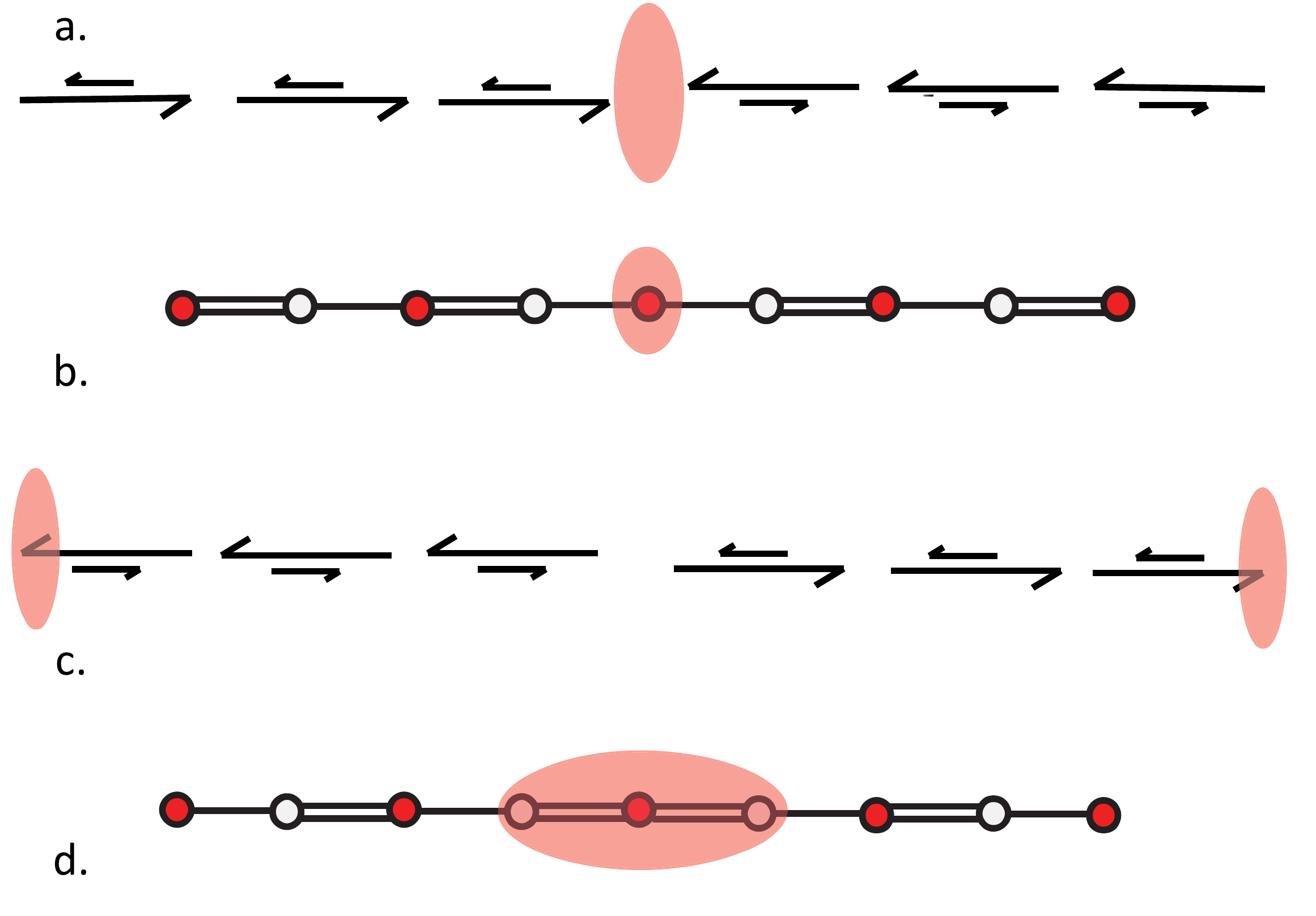}
\caption{Topological protection in a 1D out-of-equilibrium Markov state network. (a) and (c) show 1D Markov networks with an interface: the transition rates are different in the left and right subregions of the network. (b) and (d) show corresponding lattices for the SSH model  \cite{Kane2013}, with red and white sites representing two different sublattices.   (a) shows the case when the net probability flow is towards the interface leading to localization of the steady state probability density there, (b) is the corresponding SSH lattice with an energy zero mode localized on the red site at the interface. (c) is when net probability flow is away from the interface leading to localized probability density at either end of the chain. (d) is the corresponding SSH lattice with an energy zero mode localized on the white sites at the interface, and on the red sites at the ends.}
\label{fig:1}
\end{figure}

In this section, we consider the out-of-equilibrium statistical dynamics of stochastic processes described by a Markov network \cite{Schnakenberg1976}. Such descriptions are routinely used in statistical mechanics as reduced models of chemical and biophysical processes \cite{van2011stochastic}.  For equilibrium systems, the steady states and dynamics of fluctuations about it can be described in terms of energy landscapes, but such a simple description is not available for out-of-equilibrium processes. Hence, any insight into the existence and robustness of steady states of systems driven out of equilibrium by energy-consuming processes, such as those involved in biological functions, is potentially valuable. 

Here, we focus on the steady state of lower dimensional networks characterized by uniform or nearly uniform transition rates between various mesoscopic states. This is in analogy with periodically ordered lattices in electronic and meta-materials that host localized topological states at their edges\cite{Hasan2010}. We show in the following that the steady states of certain Markov networks can indeed be mapped onto the ground states of SSH like models for 1D periodic systems with topological properties\textendash the ``forward'' and ``backward'' transition rates in the master equation play the role of the hopping rates between the two sites of a unit cell in the SSH model \cite{SSH79}. 

A simple illustration of this is depicted in Fig.~\ref{fig:1}: the probability flow  in a one dimensional ($1$D) Markov chain with constant rate of  site-to-site transition rates in the bulk regions on the left and right of an interface separating them. 
We aim to show  \textendash ~ by mapping the steady state of the 1D Markov chain to the ground state of a Hamiltonian of a topologically non trivial tight binding model (Fig.~\ref{fig:1}(b))\textendash ~ that a possibly ``disordered'' interface connecting two ``bulk'' regions of the network with different transition rates can host localized topological modes depending on the transition rates in the bulk.  
Our results follow from the bulk-boundary correspondence inherent in topological systems, where the existence of localized zero modes at an edge can be predicted by studying the properties of the system in the bulk \cite{Kane2013}. 

The procedure for establishing the mapping between the stochastic process and the Hamiltonian of a topological tight binding model is distinct and more direct than previous work in which we suggested that the properties of certain out-of-equilibrium Markov states can be understood in terms of topological winding numbers \cite{Vaikunt2017}. Indeed, we explicitly provide forms of tight binding Hamiltonians whose ground states are the steady states of the out-of-equilibrium stochastic processes we consider. While the 1D network in Fig.~\ref{fig:1} is fairly trivial, the procedure outlined below can be used to construct effective tight binding Hamiltonians for more complex biophysical networks with many cycles.

We begin by recalling that the dynamics of Markovian systems can be modeled using a master equation 
\begin{equation}
\label{eq:master}
\frac{\partial {\bf P}}{\partial t}={\bf W P},
\end{equation}
where the vector, ${\bf P}$, denotes the probability of occupancy of various distinct states (nodes in a Markov network).  This is evolved in time by a transition matrix, ${\bf W}$, with elements $W_{ij}$ indicating the rate of transition from state $j$ to state $i$.
The zero right eigenvector of the master equation, $\ket{u}$, specifies the unique steady state of the dynamics \cite{van2011stochastic}. We are interested in conditions under which this steady state zero mode is localized at the interface. Formally, this requirement can be expressed as,
\begin{equation}
\label{eq:tracelabel}
\lim_{\epsilon\rightarrow 0}{\rm Tr}\frac{\epsilon\rho}{{\bf W}+\epsilon}=Tr[\rho \ket{u}\bra{1}],
\end{equation}
where $\rho$ is a diagonal matrix with elements $\rho_i=1$ for nodes $i$, that lie in the interfacial region and $\rho_i=0$ otherwise. Here, $\bra{1}$ is the zero left eigenvector of the master equation corresponding to the zero right eigenvector, $\ket{u}$. The trace in Eq.~\ref{eq:tracelabel} counts the number of zero modes of ${\bf W}$ that are localized at the interfacial region \cite{Kane2013}. This is exactly equal to $1$ if the unique steady state solution, $\ket{u}$, is localized at this interface. 
% Since the number of eigenvectors of the master equation with an eigenvalue zero can at most be one [REF], the trace in Eq.~\ref{eq:tracelabel} can at most be equal to one. 

We now show that the condition in Eq.~\ref{eq:tracelabel} can be  related to a topological quantity calculated from the master equation in the bulk network. The state to state transition matrix, ${\bf W}$, does not itself possess the symmetries usually associated with topological protection in electronic or mechanical materials \cite{Hasan2010, Huber2016}. The eigenvalue spectrum of the master equation necessarily has one zero eigenvalue with the rest of the eigenvalues being less than zero \cite{van2011stochastic}. Further, ${\bf W}$ is usually non-hermitian and can have complex eigenvalues. In this form, Eq.~\ref{eq:master} does not possess any obvious topological properties. 

In order to uncover the topological properties of the master equation, we first note that the rate of change of probability can be expressed as~\cite{Jack2009}, 
\begin{eqnarray}
\label{eq:decomposition}
\frac{\partial {\bf P}}{\partial t}&=&{\bf W_0} {\bf J} , \nonumber \\
{\bf J} &=& {\bf W_{1}} {\bf P},
\end{eqnarray}
where the first is a continuity equation that expresses the conservation of probability with ${\bf W_0}$ being a discrete representation of the divergence operator, and ${\bf J}$ is a vector of currents across each link in the network. The matrix ${\bf W}_0$ depends only on the topology of the network and not on the transition rates. The current vector, ${\bf J}$, can in turn be expressed in terms of the probability vector, ${\bf P}$, through the matrix ${\bf W}_1$, which depends on the transition rates in the network.  

We are interested in cases when the trace count as described by Eq.~\ref{eq:tracelabel} predicts the existence of localized zero modes. We will show below with concrete examples that Eq.~\ref{eq:tracelabel} can be expressed as a topological invariant by using the fact that the information about the interface between two homogeneous bulk subregions is contained in ${\bf W}_{1}$ and not ${\bf W}_{0}$.

As an illustration of these steps, we first consider the $1$D Markov chain in Fig.~\ref{fig:1}. The dynamics of the random walker can be described by the master equation in Eq.~\ref{eq:master}.  Using the above mentioned decomposition into ${\bf W}_0$ and ${\bf W}_1$, we will show that the steady state properties of the $1$D random walker map onto those of the well known SSH model \cite{SSH79, Kane2013}. The central argument is that: if ${\bf W}_1$ has a zero right eigenvector, this eigenvector is the unique right eigenvector of ${\bf W}$ and hence the steady state accessed by dynamics under ${\bf W}$. The topological properties of the zero eigenstate of ${\bf W}_1$ can be inferred by constructing the Hermitian matrix,  
\begin{equation}
{\bf H} =
\begin{pmatrix} 
 0 & {\bf W}_{1}\\
 {\bf W}_{1}^{\rm T} & 0
\end{pmatrix},
\end{equation}
and considering the trace,
\begin{equation}
\label{eq:tracelabel2}
\lim_{\epsilon\rightarrow 0}{\rm Tr}\left[{\epsilon\sigma_z\rho}\frac{1}{{\bf H} +\epsilon}\right]=\delta n,
\end{equation}
where $\sigma_{z} \equiv \begin{pmatrix} 1 & 0 \\  0 & -1 \end{pmatrix}$ is the Pauli z matrix. The trace in Eq.~\ref{eq:tracelabel2}, denoted by $\delta n$, is the difference in the number of zero modes of ${\bf W}_1$ and  ${\bf W}_1^{\rm T}$, that are localized in  the interfacial region. Since ${\bf W}_1$ is constrained by the properties of the master equation to have only $1$ zero eigenvector, the trace in Eq.~\ref{eq:tracelabel2} can only take values $\delta n \in \{-1,0,1\}$. The steady state is localized at the interface between the two bulk regions when $\delta n=1$ as shown in Fig. 1a, whereas it is localized at the opposite ends when $\delta n =-1$, as shown in Fig. 1c.

That $\delta n$ in Eq.~\ref{eq:tracelabel} can be expressed as a difference of integer topological winding numbers characteristic of the left and right bulks, $\nu_{L/R}$, is well-established as an index theorem in the context of electronic and mechanical topological lattices \cite{Kane2013}. Systems with $\delta \nu \equiv \vert \nu_{L}- \nu_{R} \vert \neq 0$ have zero modes that are topologically protected and resistant to perturbation. These arguments show that the steady state of the master equation ${\bf W}$ is localized and topologically protected whenever ${\bf W}_1$ is topologically non trivial.

The topological winding numbers for the 1D random walker can be derived by considering the transition matrix in one of the bulk regions, ${\bf W}^{\rm b}$:
\begin{equation}
{\bf W}^{\rm b} = {\bf W}^{\rm b}_{0} \cdot  {\bf W}^{\rm b}_{1}
 =\begin{pmatrix} 
 1 & -1 &   &  \\   & 1 & -1 & \\  &  & . &.  \end{pmatrix} \cdot
 \begin{pmatrix} -v &  &  & \\ w & -v &  & \\  & . & . &   \end{pmatrix},
 \label{eq:decomp} 
\end{equation}
where $w$, $v$ denote rates of forward and back hopping in the bulk. In Eq.~\ref{eq:decomp}, $ {\bf W}^{\rm b}_{0}$ is a discrete representation of a gradient operator that encodes the topology of the network while ${\bf W}^{\rm b}_{1}$ depends on the hopping rates. 
Because of the repeated nature of these matrices corresponding to the periodic symmetry of the Markov chain in real space, they can be diagonalized in a Fourier basis, $\ket{e^{i k x}}$, in the bulk. They can then be compactly represented as ${\bf W}_{0}(k) = 1- e^{-i k}$ and ${\bf W}_{1}(k) = -v + w e^{i k}$, where the lattice constant is taken to be unity. The winding number of the bulk region can be obtained from this Fourier representation. Specifically, the bulk Hermitian matrix ${\bf H}$ can also be diagonalized in a Fourier basis,  
\begin{equation}
{\bf H}(k) =
\begin{pmatrix} 
 0 & {\bf W}_{1}(k)\\
 {\bf W}_{1}^{T}(k) & 0
\end{pmatrix}
= \begin{pmatrix} 
0 & -v+ w e^{i k} \\  -v + w e^{-i k}  & 0 
\end{pmatrix} .
\label{defineH}
\end{equation}
In the form of Eq.~(\ref{defineH}), one can see that ${\bf H}$ is isomorphic to the Hamiltonian of the quantum SSH model with hopping rates, $-v$ and $w$. Using the results obtained for the SSH model, the winding number $\nu$ for the bulk phase can be calculated as \cite{Kane2013}, 
\begin{equation}
\nu =\frac{1}{2\pi \rm{i}} \int_0^{2\pi}  dk \frac{d}{dk} \ln W_1(k).
\end{equation}
% check with Suri that this is right.

\begin{figure*}[tbp]
\includegraphics[width=1\linewidth]{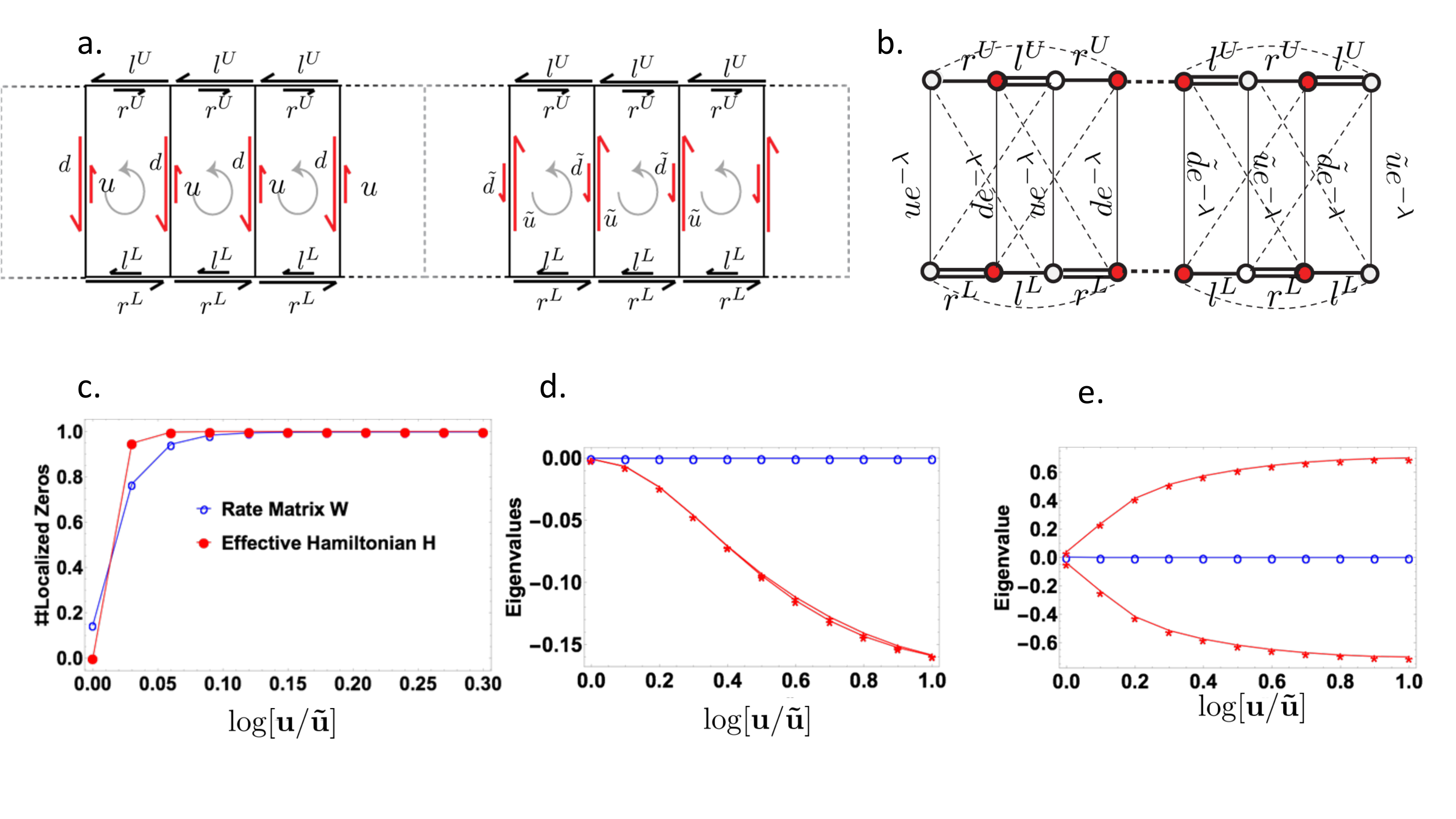}
\caption{Topological protection in a model out-of-equilibrium Markov state network (a) The ladder network with two coupled 1D Markov chains. The transition probabilities to go left, right, up or down are labeled $l$, $r$, $u$ and $d$ respectively to the left of the interface. The right bulk-like region is distinguished from the left by the different vertical transition rates: $\tilde{u}$, and $\tilde{d}$, that lead to an opposite polarization on either side of the interface. The superscripts,  $U$ and $L$, denote the upper and lower rails of the ladder. (b) The analogous tight-binding hopping model for electrons on a lattice. It involves higher order couplings between next nearest neighbors (detailed expressions in the SI). (c) The number of localized zero modes calculated from the trace count in Eq. 2 for both the master equation, ${\bf W}$ for the ladder network, and the ${\bf H}$ operator obtained from it by the construction in Eq. 4. (d) The first two  eigenvalues of ${\bf W}$ demonstrating the eigenvalue gap closes when the transition rates are equal on the left and right subregions.  (e) The corresponding first two eigenvalues of ${\bf H}$. The plots in (c),(d),(e) are shown as a function of the inverse localization length which is related to the ratio of vertical rates on either side of the interface: $u/\tilde{u} = \tilde{v}/v$. }
\label{fig:2}
\end{figure*}

Thus we have mapped the zero modes of the 1D Markov chain directly to those of a corresponding SSH model with the backward and forward transition rates playing the role of the two hopping parameters in the tight-binding model of the Hamiltonian.  The polarization generated in a topologically nontrivial SSH model is simply related to current generated by the bias in the random walk model. In this sense, dissipation in the bulk of the random walk model plays a crucial role in the generation of localized states. In analogy with the SSH model, two connected chains with opposite polarizations of probability flux will naturally lead to a  accumulation of probability at the interface as the system approaches steady state.

The Markov state model in Fig.~\ref{fig:1} does not possess multiple cycles. Such cycles can allow for feedback at the cost of energy dissipation and are features common to Markov state representations of many out-of-equilibrium biophysical processes \cite{Tu2008}. To derive our results in the context of out-of-equilibrium stochastic models relevant for biological processes, we consider the minimal Markov state model shown in Fig. 2a. that was introduced in Ref.~\cite{Vaikunt2017}. This ladder-like Markov network possesses two horizontal rails with transition rates, $l^{U}$,$ r^{U}$, $l^{L}$ and $r^{L}$ denoting the leftward and rightward transition rates in the upper and lower rails. There is also an upward and downward transition probability along each vertical rung of the ladder denoted by $u$ and $d$ respectively. 

 The Markov state model is composed of two translationally invariant \textit{bulk} like regions with an \textit{interface} connecting them. Specifically, the rates of transitions in the \textit{bulk} regions do not depend on the position along the horizontal axis. The rates in the interfacial region interpolate between the two bulk regions. The transition rates in the right bulk region are denoted by the $\,\tilde{}\,$ symbol to distinguish them from those on the left. As discussed in Ref.~\cite{Vaikunt2017}, the spatial connectivity and structure of this Markov state network resembles that of networks routinely used to study adaptation~\cite{Lan2012}, kinetic proofreading~\cite{Murugan2012,Murugan2016}, and cell signal sensing~\cite{Mehta2012}. These and other Markov state representations of biophysical processes can often be decomposed into \textit{bulk} like subgraphs stitched together by interfaces as indicated in Fig. 2a. The subgraphs are formed by finite periodic replication of a particular module or motif.

Since we are mainly interested in networks of the form in Fig.~\ref{fig:2}, which possess translational symmetry along one (horizontal) axis and the interface spans the other (vertical) axis, we decompose rate matrix of this system as 
\begin{equation}
\label{eq:defineWdecompose}
{\bf W}={\bf W}_0^x {\bf W}_1^x + {\bf W}_0^y {\bf W}_1^y,
\end{equation}
where ${\bf W}_{0/1}^{x/y}$ are square matrices and are discrete representations of the continuity operator in the (horizontal) $x$ and (vertical) $y$ directions. 
Since the interface spans the vertical axis, we choose the decomposition, 
\begin{equation}
\label{eq:defineWdecompose1}
{\bf W}={\bf W}_0^x ({\bf W}_1^x + {{\bf W}_0^x}^{-1} {\bf W}_0^y {\bf W}_1^y)\equiv {\bf W}_0^x \tilde{{\bf W}_1^x}.
\end{equation}
These arguments imply that any master equation rate matrix ${\bf W}$ can be factorized as, ${\bf W}={\bf W}_0^x \tilde {\bf W}_1^x$. Again, the crucial point in this decomposition is that ${\bf W}_0^x$ does not depend on the transition rates and possesses no interfaces. Using this property and arguments similar to the case of 1D random walk model, one can show that {\bf W} has topologically protected modes whenever the following Hermitian operator constructed with $\tilde {\bf W}_1^x$ is topologically non trivial, 
\begin{equation}
\label{eq:defineH}
{\bf H} =\left( \begin{array}{cc}
0 & \tilde {\bf W}_1^x\\
(\tilde {\bf W}_1^x)^{\rm T} & 0  \end{array} \right).
\end{equation} 
Like the master equation rate matrix, the Hamiltonian ${\bf H}$ is also composed of two bulk phases connected by interfaces. A schematic of the connections in the effective Hamiltonian is provided in Fig.~\ref{fig:2}. We provide the detailed derivation and explicit expression for ${\bf H}$ in the Supplementary information.

The polarization implicit in the Hamiltonian ${\bf H}$ reflects the currents generated by the master equation rate matrix ${\bf W}$. For instance, the effective Hamiltonian we have depicted in Fig.~\ref{fig:2}b, is composed of two horizontal rails each with a lattice structure that resembles that of the SSH model. The two rails can potentially have polarizations with the same magnitude but opposite signs. This choice corresponds to the case  $ l^{U}=r^{L}$ and $ r^{U}=l^{L}$. In such cases, the effective Hamiltonian can have a net polarization  only when the vertical links connecting the two rails break symmetry. Indeed, this condition reflects the requirement that the symmetry between the two rails of the ladder in the master equation rate matrix be broken, $u \neq d$, for a current along the horizontal axis of the bulk networks. In this sense, topological protection depends crucially on the currents generated by the dissipative fluxes in the master equation rate matrix. In the context of the models considered here, systems without dissipative fluxes in their bulks cannot support a topologically protected mode.
In Fig.~\ref{fig:2}c, we show the numerically computed count of localized zero eigenmodes of ${\bf W}$ using Eq.~\ref{eq:tracelabel}, and compare it to that of the effective matrix, ${\bf H}$. We also show the gap in the eigenvalue spectrum for ${\bf W}$ and ${\bf H}$ in Figs.~\ref{fig:2}c \& d respectively, by numerically computing the first two eigenvalues as the hopping rates are varied.  Such a gap between the steady state and subsequent eigenvalues in the eigenvalue spectrum make the steady state robust against random fluctuations. We indeed find that as long as  the Hamiltonian $H$ has a winding number mismatch that supports a localized zero mode at the interface, ${\bf W}$ is also guaranteed to have a zero mode localized to the interface.  

This is the main result of the first part of the paper. It establishes that the topological properties of ${\bf W}$ can be computed by simply constructing winding numbers for the matrix, $\tilde{\bf W}_1^x$. This mapping allows us to infer the properties of a stochastic out-of-equilibrium system in contact with thermal reservoirs in terms of the topological properties of an isolated quantum mechanical system  described by the Hamiltonian ${\bf H}$. The Hamiltonian reflects the polarization of the bulk master equations that support a current.

\section{Topological protection in many body systems}

%In this part of the paper, we address the critical question of whether ideas from topological protection can be applied to active soft matter physics. 
There is broad interest in dissipative, steady-state structures (such as ordered phases and topological defects therein) that emerge in collections of particles driven away from equilibrium in both synthetic \cite{Cates2015} and biological contexts \cite{RMP13, Kruse2005}.  Such steady states, if topologically protected, are likely to be robust against disorder and can be categorized into distinct topological classes -- thus guiding applications that involve organization away from equilibrium \cite{Whitelam2012,Nguyen2016}.

The results of the previous sections are, however, specific to effectively single particle Markov state processes. A natural question is whether similar statements about topological modes can be made for collections of many interacting particles out of equilibrium.
Indeed, an effective Markov state representation of the dynamics of a many particle interacting system cannot be simplified in the same way as the models in Figs~\ref{fig:1} \& ~\ref{fig:2}. While a full microscopic description of a many-particle interacting system is in general intractable without doing detailed simulations, the relaxation of such a system towards equilibrium or a steady state can be described in terms of a few slowly decaying collective modes with long wavelengths.  Such a hydrodynamic description in terms of long wavelength collective modes differs fundamentally from the electronic or mechanical materials with periodic order for which topological physics is typically demonstrated \cite{Hasan2010, Kane2013}.

In this section, we show that the equations of motion derived for a set of $m$ hydrodynamic fields, 
$\{ {\bf X_{i}} ({\bf x},t) \}$, each corresponding to a conserved quantity, can have steady state solutions that are topologically protected. We focus on purely dissipative hydrodynamics, where the principles of linear irreversible thermodynamics derived in the seminal works of Prigogine, Onsager, deGroot and Mazur \cite{deGroot1969} can be used to write the equations of motion as,
\begin{equation}
\frac{d {\bf Y}}{dt} = - {\bf M} {\bf Y},
\end{equation}
where ${\bf Y} = \{{\bf X_{1}},...,{\bf X_{m}} \}$ is the set of all the $m$ hydrodynamic variables in a given problem, and the matrix ${\bf M}$ contains the dissipative couplings between the various hydrodynamic variables.  Since we consider purely dissipative processes, an excited collection of particles should relax monotonically in time towards the steady state which corresponds to minimal or no dissipation. This constrains the eigenvalues of ${\bf M}$, which controls the dynamics of the variables $ {\bf X_{i}} ({\bf x},t)$, to be positive.   This property allows us to decompose a suitably symmetrized form of ${\bf M}$ as a matrix product ${\bf D} {\bf D}^{\rm T}$. We can then extract its topological properties using suitable index theorems for the \textit{square root} operator given in Ref.~\cite{Kane2013}: ${\bf S} = \begin{pmatrix}  0 & {\bf D} \\ {\bf D}^{\rm T} & 0 \end{pmatrix}$, that possesses a symmetry, $ \{\sigma_{z},{\bf S}\} =0$, that leads to positive and negative pairs of eigenvalues, and  $S^{2}$ has the same eigenvalue spectrum as the original dynamical operator, ${\bf M}$, including its zeros.

With this background in mind, we now demonstrate topologically protected flow in a simple fluid setup as a prelude to application to more complex chiral, active fluids \cite{Tsai2005, Strempel2013}.  We consider a thin layer (quasi 2D in the $xy-$plane) of viscous, incompressible, fluid on a solid substrate and bound between two vertical plates(positioned in the $yz-$ plane). Moving one of these bounding plates with velocity $v_{0}$ induces a flow in the $y-$direction.  The hydrodynamic equation for the velocity of the fluid is determined by momentum conservation which includes the dissipative processes in both the bulk of the fluid and at its surface in contact with the substrate which acts as a ``momentum sink''. In the linear response regime, the dissipation potential governing the flow in this setup is, $\mathcal{R} = \int d{\mathbf x} \, (\eta (\nabla \mathbf{v})^{2} + \gamma \mathbf{v}^{2})$, where $\eta$ is the fluid viscosity and $\gamma$, the substrate friction coefficient. The equation of motion obtained by extremizing the dissipation functional, $\mathcal{R}$, is: $\rho Dv/Dt = (\nabla^{2} - \lambda^{-2}) v$, where $-(\nabla^{2} - \lambda^{-2})$ is the hydrodynamic operator, ${\bf M}$, for this setup, and $\lambda \equiv \sqrt{\eta/\gamma}$ is a friction length.  The resulting steady state velocity profile is exponentially localized near the moving plate: $v_{y}(x) = v_{0} e^{-x/\lambda}$ assuming no slip boundary conditions.  

Although, we need appropriate boundary conditions (here, no slipping at the plate surface) to obtain the exact spatial profile of the fluid velocity,  topological protection \textemdash which we demonstrate in the next paragraph \textemdash implies that the fluid velocity is localized near a moving plate near the plate irrespective of the details of the boundary conditions, as long there is a finite dissipative coupling ($\gamma >0 $) with the substrate.

The simple linear operator, $ \nabla^{2} - \lambda^{-2} $, which occurs generically in dissipative hydrodynamics, can be characterized by a topological index that is related via an index theorem to the localized zero mode of the operator, $i.e.$ the steady state solution of the hydrodynamics.  To see this explicitly, we construct the \textit{square root}, $-(\nabla^{2} - \lambda^{-2}) = {\rm D}_{0}{\rm D}_{0}^{T}$, of the linear hydrodynamic operator. This may be represented by a finite ($N \times N$) matrix by discretizing the Laplacian operator on a $1D$ mesh in real space with a lattice constant, $h$:
\begin{eqnarray}
 {\bf M} &=& -\begin{pmatrix} 
 -2 h^{-2}- \lambda^{-2} & h^{-2} & 0 &  & \\ h^{-2} & -2 h^{-2}- \lambda^{-2} & h^{-2} &  & \\  & . & . & .& \\  &  & . & . & .\end{pmatrix} \nonumber \\
 &=& \begin{pmatrix} 
 v & -w &  & &  \\   & v & -w &  & \\  &  & . & . &\\  &  &  & . &. \\ & &  &  & v \end{pmatrix} \cdot
 \begin{pmatrix} v &  & & &\\  -w & v &  &  &\\  & . & . &  &\\  &  & . &.  &  \\ &  &  & -w & v \end{pmatrix},
 \nonumber \\
\label{squareroot:1}
\end{eqnarray}
where the $D_{0}$ matrix is seen to be identical to the SSH Hamiltonian, with $v,w = \frac{1}{2} \big( \sqrt{\lambda^{-2} + 4 h^{-2}} \mp \lambda^{-1} \big)$.  As shown in the case of an SSH chain \cite{SSH79}, the interface with vacuum (in this case, a hard wall) can host a localized edge mode (in this case, of the velocity at steady state) when the bulk is characterized by a finite topological index, as it happens for $w > v$. This is satisfied in the choice of decomposition above irrespective of the fineness of the mesh, \emph{i.e.} the value of $h$.  The only exception to this is for the case $\lambda^{-1} = 0$ \emph{i.e.} there is no friction from the substrate, when the above construction results in $v=w$, the topologically trivial case corresponding to a closing in the gap of the eigenvalue spectrum of $(\nabla^{2} - \lambda^{-2})$. The details of the calculation and an argument connecting friction to an effective polarization are presented in the SI. We note that while the lattice discretization $h$ was introduced to construct the decomposition in Eq.~\ref{squareroot:1}, the topological index is independent of it. Finally, the decomposition of the linear operator is valid even with $w$, $v$ interchanged. While such a choice leads to a trivial topological state, it doesn't invalidate our conclusions. Indeed, the topological state of the SSH model is similarly changed when the unit cell is shifted by one lattice unit. For the hydrodynamic equations to have a topological mode, we simply require that there exists a decomposition that leads to a non zero topological index. 

As a nontrivial application of this notion of topological protection to a many-particle system, we focus on collections of actively rotating particles (rotors) with an intrinsic ``spin'' angular momentum degree of freedom. The collective dynamics of such particles are more complex and relatively less understood than those that involve linear self-propulsion \cite{Cates2015} though there is now a wide range of experiments where such active angular momentum injection in the bulk can be realized in a controlled manner (see Refs. \cite{vanZuiden16, Nguyen2014}).  Examples include liquid crystals in a rotating magnetic field \cite{Migler1991}, shaken chiral grains \cite{Tsai2005} and light-powered colloids \cite{Maggi2015}. Instances of naturally occurring flowing matter with actively generated rotation by  molecular motors range from the rotational beating of flagella of swimming bacteria to active torque generation in the cellular cytoskeleton \cite{Strempel2013}. Such molecular torques are potentially biologically significant -- and have been implicated in the streaming chiral flows in the actin cortex that lead to left-right symmetry breaking of a developing organism during morphogenesis \cite{Naganathan2014}, for example.  Boundary modes where the rotors circulate around the edge of a container have been seen in Ref.~\cite{Tsai2005} and more recently in simulations in Ref.~\cite{vanZuiden16} but connections to topological protection, if any, have not been explored.  We now demonstrate that active spinners in confined geometry can indeed support topologically protected localized edge modes. These modes allow the system to robustly localize flows and transport to boundaries and provide a route to the breaking of chiral symmetry.  

\begin{figure*}[tbp]
%\begin{center}
\includegraphics[width = 6 in]{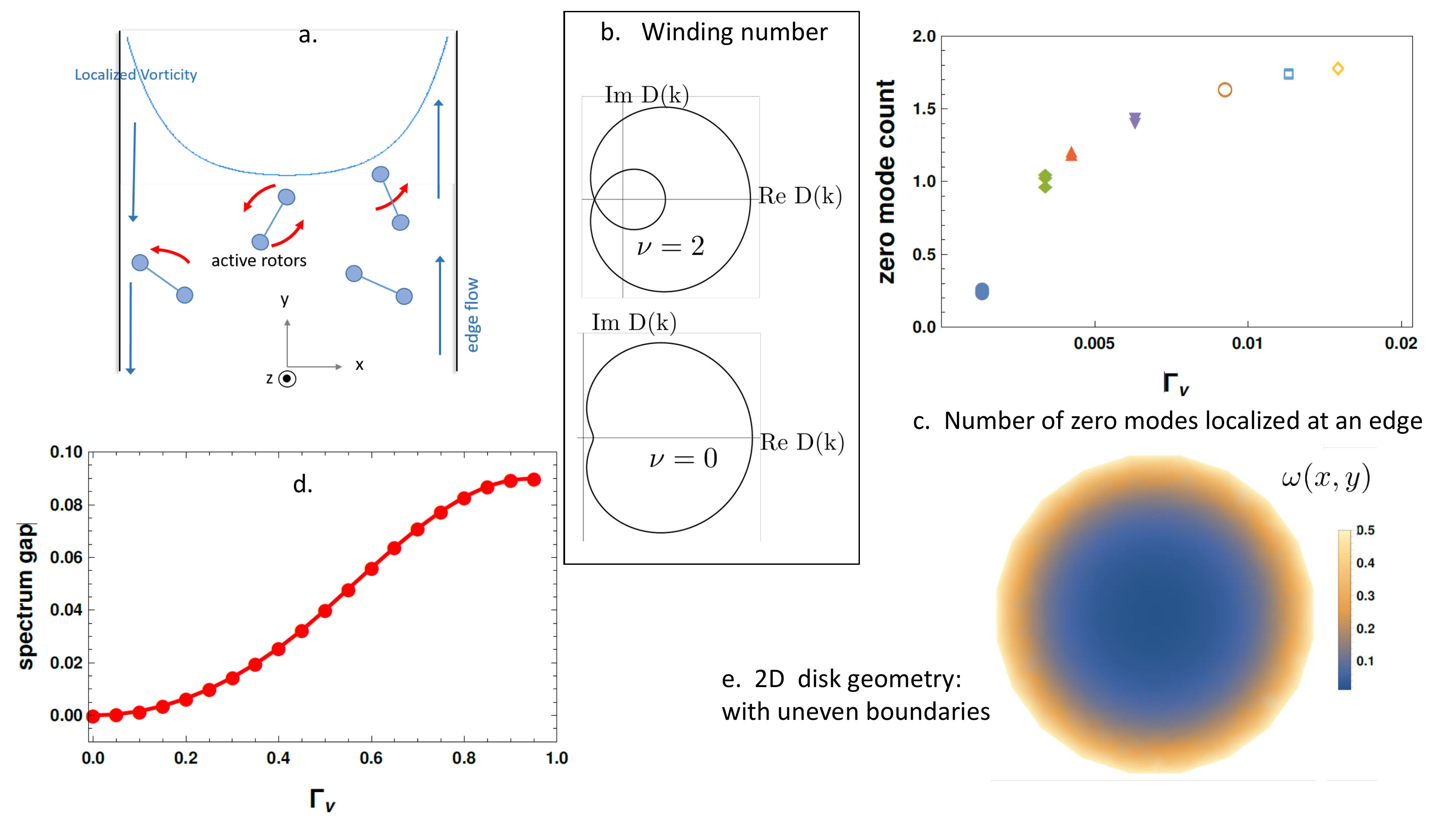}
 \caption{ {\bf Topological protection of hydrodynamic boundary flows }  a. Model of actively spinning rotors confined between two plane surfaces 
 b. Plots showing winding number based on the construction of Eq.(16).  c. Count of zero modes localized at an edge as a function of substrate friction, $\Gamma_{v}$, for 
 $10 \%$ disorder in material parameters. d. Eigenvalue gap as a function of substrate friction, $\Gamma_{v}$ showing gap closing when there is no substrate friction.  e. Localized vorticity at the boundary in 2D confined disk geometry with uneven boundaries. }
\label{fig:3}
 \end{figure*}

The dynamics of a collection of actively spinning rotors can be described by a coarse-grained hydrodynamic theory formulated in terms conservation laws and constitutive relations derived using the general principles of irreversible thermodynamics \cite{deGroot1969}. The key hydrodynamic variables in the theory are the fields of intrinsic rotation rate (or ``spin'' angular velocity), $\Omega({\bf x, t})$, and the linear flow velocity of the rotors, ${\bf v}({\bf x},t)$, corresponding to the conservation of angular momentum and linear momentum respectively.  We assume incompressibility based on the observation that the density of rotors remains nearly uniform \cite{vanZuiden16}. A crucial feature of these phenomenological  equations of motion is the dissipative coupling between the angular velocity, and the vorticity, $ \omega = (\nabla \times {\bf v}({\bf x}))_{z}/2$ through the rotational strain rate, $\Omega - \omega$, which is the generalized force corresponding to a thermodynamic flux: the antisymmetric component of shear stress  induced by the relative rotation of adjacent fluid elements \cite{Furthauer2012,Mandal2017,Klymko2017}. The antisymmetric stress, is responsible for the dissipative transfer of angular momentum from the intrinsic spin degree of freedom to the linear bulk vorticity \cite{Furthauer2012,Mandal2017,Klymko2017}, and is always present unless the spin and vorticity become equal. 
%reword:

The hydrodynamic equations corresponding to the complex rotor systems can be inferred using the principles of linear irreversible thermodyamics \cite{deGroot1969} from a dissipation functional, 
\begin{equation}
\mathcal{R} = \int d{\bf x} \big[\eta (\nabla {\bf v})^{2} + D_{\Omega} (\nabla \Omega)^{2}  + \Gamma (\Omega - \omega)^{2} + \Gamma_{v} v^{2} + \Gamma_{\Omega} \Omega^{2} \big],
\label{entropy_prod}
\end{equation}
%(when substrate friction is turned off),
derived from the proportionality of the thermodynamic forces and fluxes through the dissipative phenomenological coefficients:  the spin-vorticity coupling, $\Gamma$, the rotational diffusion constant, $D_{\Omega} $, the viscosity of the medium, $\eta$ and the substrate friction coefficients for angular and linear velocity: $\Gamma_{\Omega}$ and $\Gamma_{v}$.  These dissipative processes that contribute to the net rate of entropy production in Eq.~(\ref{entropy_prod}), together with the uniform active torque, $\tau$, driving each rotor, determine the dynamics at steady state. 

The coupled dynamical equations for the spin angular velocity (redefined as a difference from its bulk steady state value:  $\Omega({\bf x}) \rightarrow  \Omega({\bf x}) - \Omega_{b}$), and vorticity, $\omega = (1/2)(\nabla \times {\bf v})_{z} $, are derived by extremizing the above dissipation functional, $\mathcal{R}$, as:
\begin{equation}
\label{eq:defineH3}
\frac{\partial}{\partial t}  \begin{pmatrix} \Omega \\ \omega \end{pmatrix}=
D_{\Omega}\begin{pmatrix} 
 \nabla^2 - \lambda_{\Omega}^{-2} & a\\
 -b \nabla^{2} &  \nabla^2 -\lambda_{\omega}^{-2}
\end{pmatrix}
\begin{pmatrix} \Omega \\ \omega \end{pmatrix},
\end{equation} 
where the decay length scales, $\lambda_{\Omega}^{-2} = (\Gamma + \Gamma_{\Omega})/D_{\Omega}$,
and, $ \lambda_{\omega}^{-2} =  \Gamma_{v}/(\eta + \Gamma)$, and coupling coefficients, $a = \Gamma/D_{\Omega}$, $b = \Gamma/(\eta + \Gamma)$, are defined in terms of relevant dissipative parameters \cite{Tsai2005}.  This is the bulk steady state to which the collection of spinning rotors decays under the action of an active intrinsic torque, $\tau$, and in the presence of substrate friction, $\Gamma_{\Omega}$: $\Omega_{b} = \tau/(\Gamma + \Gamma_{\Omega})$ \cite{Tsai2005, vanZuiden16}. In a confined geometry, the rotors are prevented from rotating freely at the walls, which induces a spatial profile in the angular velocity, and therefore in the vorticity, at steady state \cite{Tsai2005, vanZuiden16}.

\begin{figure*}[tbp]

\includegraphics[width=1\linewidth]{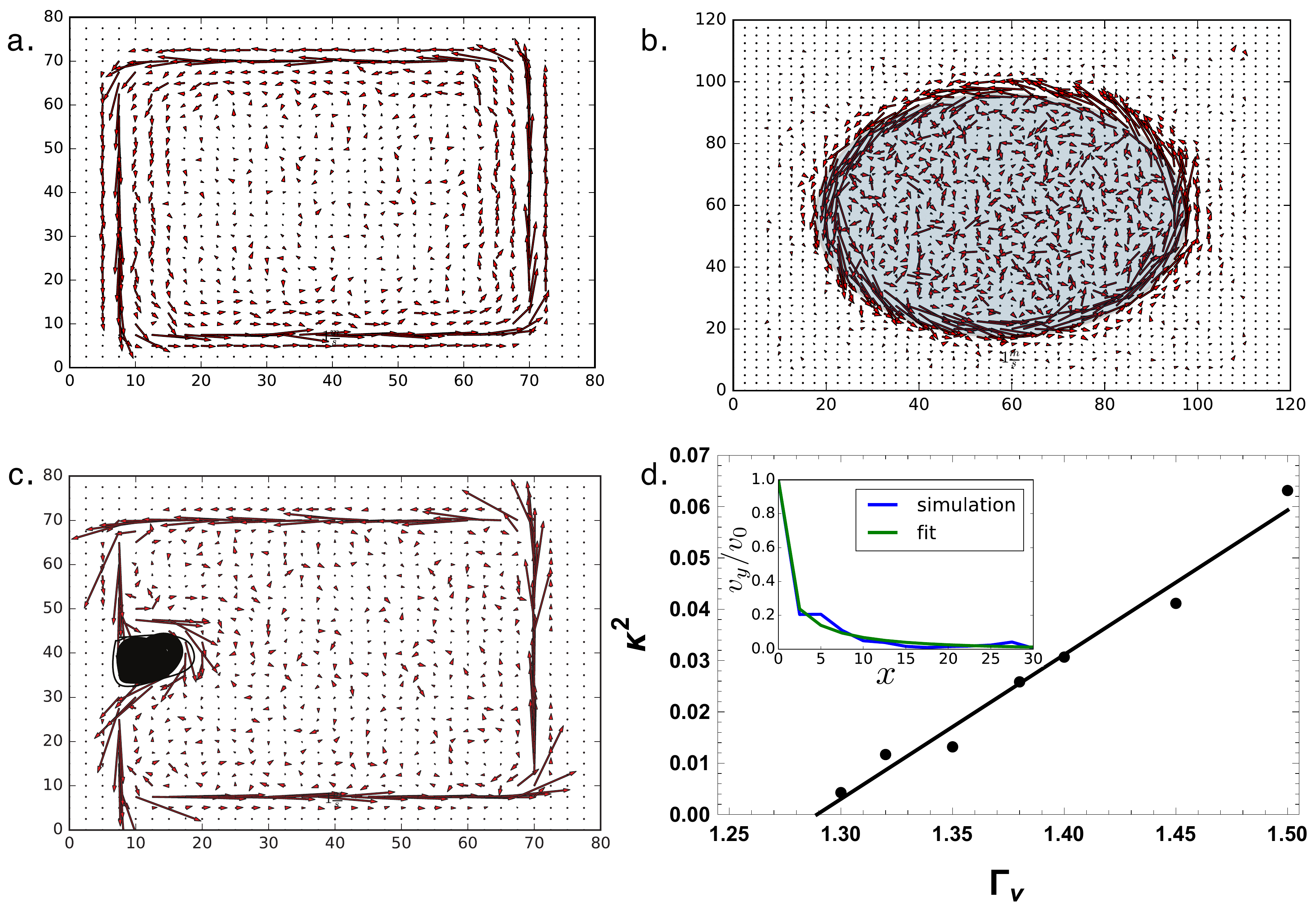}
%\begin{figure}
%\includegraphics[width=3.5 in]{fig4.pdf}
\caption{{\bf Edge flows in a collection of driven rotors
confined by wall potential.} 
Simulated edge flows in slab geometry: a. unperturbed boundaries, b. robust flow around an obstacle demonstrating lack of backscattering, c.rotors are driven in a finite subregion (shaded) of the simulation box, illustrating flow localized at the interface of the ``active'' and ``passive'' regions.  The size of the arrows is proportional to the magnitude of the average velocity and the direction of the arrows indicates the
direction of the flow.  The equations of motion used to simulate the rotors are described in the main text. The
bulk dynamics of such rotors is well described by a hydrodynamic description derived from irreversible thermodynamics. d. shows increased localization as the substrate friction, $\Gamma_{v}$ is increased.  $\kappa$ is an  effective  inverse localization length obtained from fitting the simulated velocity profile (shown in the inset, where the blue curve is the simulated velocity profile, and the yellow curve is the fit function) to an effective exponential profile. The inset axes are normalized by the box size $L$ and the maximum edge velocity, $v_{0}$.}
\label{fig:4}
\end{figure*}

We now demonstrate that Eq.~(\ref{eq:defineH3}) does indeed lead to topologically protected zero modes corresponding to its steady state solutions. The details of the derivation are given in the SI. While the full matrix ${\bf M}$  describing the rotor dynamics can be decomposed to reveal its topological properties (as we do numerically in Fig.3), the calculation of a topological index simplifies considerably  at steady state. Briefly, we integrate out the vorticity field to obtain a linear hydrodynamic operator for spin angular velocity, ${\bf L}_{s}$, at steady state. This can be discretized in real space and decomposed as $ {\bf L}_{s} =  {\bf D}_{s} {\bf D}_{s}^{\rm T}$. The resulting $ {\bf  D}_{s}$ is a sparse matrix with elements,  $({\bf D}_{s})_{i,j} = v \delta_{i,j} + w_{1} \delta_{i-1,j} + w_{2} \delta_{i-2,j}$,  that are non-zero only in the diagonal and two nearest off-diagonals.  This allows us to define a matrix, ${\bf S}_{s}$, which is represented in Fourier basis (with appropriately defined periodic boundary conditions) as,
\begin{eqnarray}
\label{eq:sshmatrix}
{\bf S_{s}} &=&
\begin{pmatrix} 
 0 & {\bf D}_{s}(k)\\
{\bf D}_{s}^{\rm T}(k) & 0 
\end{pmatrix}  \\
&  =&
\begin{pmatrix} 
0 & v+ w_{1} e^{i k} + w_{2} e^{ 2 i k} \\  v + w_{1} e^{-i k} + w_{2} e^{- 2 i k} & 0 
\end{pmatrix},\nonumber
\end{eqnarray}
%L(k) = v+ w e^{i k}, L(-k)= v - w e^{-i k}
 such that ${\bf S_{s}}^{2}$ and ${\bf L}_{s}$ have the same eigenvalues.  ${\bf S}_{s}$ corresponds to an SSH like model for 1D topological insulators with both nearest and next nearest neighbor hopping. The topological properties of this model can be characterized in terms of a winding number  \cite{Kane2013}, 
 \begin{equation}
\nu = \frac{1}{2 \pi i} \int dk \frac{d} {dk} \ln (D_{s}(k)).
 \end{equation}
 
Although we consider rotors confined in a 2D geometry, we  discretize the rotor equations along one direction transverse to the boundary (and therefore, to the edge currents), leading to a winding number \cite{Ryu2002}. We again stress that the winding number is not a function of the lattice discretization.

 In the presence of finite coupling between spin and vorticity, the gap in the bulk spectrum of ${\bf L}_{s}$ closes only if $\lambda_{\omega}^{-2} = 0$, \emph{i.e.} if friction with the substrate vanishes, $\Gamma_{v} =0$ (Fig. 3d).  This corresponds to $v + w_{2} = w_{1}$, \emph{i.e.} winding number of zero (see SI for the winding number calculation).  In general, in the presence of finite friction, the winding number is $\nu = 2$ corresponding to two localized modes. The exact spatial profile of the spin velocity and vorticity is a linear superposition of these independent modes and depends on the specific boundary conditions imposed in a confined geometry \cite{Tsai2005}.
 
%  Thus, in the presence of finite substrate friction,  exponentially localized edge flows are topologically guaranteed in a collection of spinning rotors.  
% {eq:defineH3}
 
In Fig.~ \ref{fig:3} we plot the eigenvalues of the discretized matrix for a slab geometry. The eigenvalues are real and non-positive. We calculate numerically the trace in Eq.~(\ref{eq:tracelabel}) for the full rotor hydrodynamic operator defined in Eq.~(\ref{eq:defineH3}), to show that there are indeed two localized zero modes in the presence of friction and finite spin vorticity coupling. This corresponds to the winding number, $\nu =2$, calculated for the rotors at steady state. Further, we observe that in the presence of friction and finite spin vorticity coupling, there is a gap in the spectrum of the eigenvalues between the zero eigenvalue and subsequent eigenvalues. As discussed in other contexts~\cite{Kane2013}, such a gap protects the steady state zero eigenvalue solution against perturbations. The gap and the localized modes mirror features encountered during the analysis of the master equation. Like in the master equation analysis, the hydrodynamic equations of motion in Eq.~\ref{eq:defineH3} are dissipative, non-Hermitian and their topological properties if any, are not immediately apparent.

As a further test of these ideas, we set up molecular dynamics simulations of a system composed of driven rotors confined in a box. It has been well established that the hydrodynamic equations in Eq.~\ref{eq:defineH3} are good descriptors of the dynamics of such rotor systems \cite{Tsai2005, vanZuiden16}. The equations of motion of individual rotors in our molecular dynamics simulations are:
\begin{eqnarray}
m \ddot{{\bf r}_{i}} = -\Gamma_v {\bf r}_{i} - \partial_{r_{i}} \sum_{i\neq j} V({\bf r}_{j} - {\bf r}_{i}, \theta_{i}, \theta_{j}) + D_1 \eta_1(t) , \nonumber \\
I \ddot{\theta} = \tau - \gamma_{\Omega} \dot{\theta}_{i} -  \partial_{\theta_{i}} \sum_{i\neq j} \mathcal{V}({\bf r}_{j} - {\bf r}_{i}, \theta_{i}, \theta_{j}) +D_2 \eta_2(t),
\label{simeq}
\end{eqnarray}
where $\Gamma_v$ is the substrate friction, $\gamma_{\Omega}$ is the friction imposed on the $\theta$ degree of freedom, $\tau$ is the applied torque, $\eta_1(t), \eta_2(t)$ are delta function correlated Gaussian random variables, and the interaction potentials $V$ are derived from the Yukawa pair potential \cite{vanZuiden16}. The detailed parameters used for the simulations are described in Methods.  

In order to check for the existence of localized vorticity predicted by the theory, we computed the time averaged velocities of the center of mass of the rotors. As is clear from Fig.~\ref{fig:4}a , this system supports localized edge flows.  This flow is immune to backscattering in the presence of obstacles  and persists even if the edges of the box used in the simulation are jagged (as depicted in Fig.~\ref{fig:4}b). Further, when the rotors are actively driven only in a partial region of the simulation box (Fig.~\ref{fig:4}c), the flow is localized at the boundary of this region of driving. 

According to a solution of the hydrodynamic equations with a specific choice of boundary conditions~\cite{vanZuiden16} (SI), one expects the inverse localization length $\kappa$ to scale like $\kappa^2 \propto \Gamma_v$.  This solution, and the theoretical analysis above, predicts a loss of topological protection at zero substrate friction. In agreement with these predictions, the boundary flow velocity continues to be robustly localized  as substrate friction $\Gamma_v$ is reduced until very small values of friction, when the flow profile turns noisy (Fig.~\ref{fig:4}d). The loss of topological protection in a finite system at non-zero friction can be interpreted as a signature of the narrowing of the associated gap in the eigenvalue spectrum of flow solutions in relation to fluctuations as friction is decreased, resulting eventually in an unstable steady state. Similar features are apparent as topological protection is lost in finite size electronic and mechanical topological insulator models \cite{Lubensky2016}. 
These observations taken together constitute a demonstration of the robustness of edge flows to boundary conditions. 

We have shown here that hydrodynamic equations describing the dynamics of rotors can  be characterized by a topological index.  This quantity is calculated for the bulk of the flow and guarantees the existence of flows localized at the edge of a confined fluid as a consequence of the  principle of bulk-boundary correspondence well-known in topological physics  \cite{Hasan2010, Kane2013}. This conclusion is reached without explicitly solving hydrodynamic equations which require one to assume hydrodynamic boundary conditions that are not always intuitive.  For example, the direction of the edge flow seen in these active rotors is not obvious \emph{a priori}.  Although the flow velocity obtained from simulations has been fitted to the flow profile corresponding to zero tangential stress from the confining wall \cite{vanZuiden16}, we show in the SI that a no-slip boundary condition at the wall also leads to a localized velocity profile. This shows that the solutions guaranteed by topological considerations are independent of the exact boundary conditions for a given hydrodynamic problem.

Using two disparate  examples of complex dynamical systems that are out of equilibrium, we have shown in this paper that topologically protected states can arise in principle in a variety of dissipative systems: both stochastic networks and active flows. Unlike other systems for which topological protection has been previously explored, such as mechanical lattices \cite{Kane2013} or propagating sound modes in hydrodynamic equations~\cite{Souslov2016,Shankar2017}, dissipation is key to the phenomena considered here. In both cases considered, the topological properties are not readily apparent from the structure of the relevant operators that govern their dynamics. We reveal their topological properties by decomposing them suitably in order to map the properties of their steady state solutions to the zero energy states of 1D Hamiltonian models that can be characterized by a topological index. Our work indicates that both single and many particle dynamics with interactions at both microscopic and macroscopic scales can result in topological modes that are localized at boundaries. Further, our results also suggest that the chiral edge flows ubiquitously seen in synthetic and biological matter are  potentially robust, topologically protected modes. The robustness in this case is explained using broadly applicable ideas from linear irreversible thermodynamics. This has applications for self-assembly of metamaterials as well as guarantee robust localized flows of both information and matter in biology. 
 
\section{Acknowledgments}
We gratefully acknowledge very useful discussions with William Irvine, Sid Nagel and Tom Witten. K D. and S.V. were funded by NSF DMR-MRSEC 1420709. SV also acknowledges funding from the University of Chicago and the Army Research Office under grant number W911NF-16-1-0415. KKM acknowledges support from a National Institutes of Health Grant R01-GM110066. He is also supported by Director, Office of Science, Office of Basic Energy Sciences, Chemical Sciences Division, of the U.S. Department of Energy under contract No. DE- AC02-05CH11231

\section{Methods}
The molecular dynamics simulations were performed by evolving the Langevin equation with Euler dynamics. The code is available upon request. The rotors in our simulations were composite particles composed of three equally spaced point particles arranged along a line. The particles interact according to a Yukawa potential with a decay length $2d$ where $d$ is the spacing between the particles in the rotor. This force field was used to derive the interatomic forces and torques on the rotor. 
We set $\Gamma_v=\gamma_\Omega$ as the substrate friction. The simulations in Fig 4 a, c, d  were performed with $N=160$ particles in a two dimensional square box with length $L=80$. The size of the rotors is $d=2.5$. The charge on the rotors (for the Yukawa potential) was set to $q=2.5$. The simulations in Fig 4 b. were performed with $N=362$ particles in a box with size $L=120$. The applied torque $\tau$ in all the simulations was $\tau=10$. The variance of the Gaussian noise in simulations was set using $D_1=1$ and $D_2=0.5$.  
\bibliographystyle{apsrev4-1}
%\bibliography{references}

\begin{thebibliography}{44}%
\makeatletter
\providecommand \@ifxundefined [1]{%
 \@ifx{#1\undefined}
}%
\providecommand \@ifnum [1]{%
 \ifnum #1\expandafter \@firstoftwo
 \else \expandafter \@secondoftwo
 \fi
}%
\providecommand \@ifx [1]{%
 \ifx #1\expandafter \@firstoftwo
 \else \expandafter \@secondoftwo
 \fi
}%
\providecommand \natexlab [1]{#1}%
\providecommand \enquote  [1]{``#1''}%
\providecommand \bibnamefont  [1]{#1}%
\providecommand \bibfnamefont [1]{#1}%
\providecommand \citenamefont [1]{#1}%
\providecommand \href@noop [0]{\@secondoftwo}%
\providecommand \href [0]{\begingroup \@sanitize@url \@href}%
\providecommand \@href[1]{\@@startlink{#1}\@@href}%
\providecommand \@@href[1]{\endgroup#1\@@endlink}%
\providecommand \@sanitize@url [0]{\catcode `\\12\catcode `\$12\catcode
  `\&12\catcode `\#12\catcode `\^12\catcode `\_12\catcode `\%12\relax}%
\providecommand \@@startlink[1]{}%
\providecommand \@@endlink[0]{}%
\providecommand \url  [0]{\begingroup\@sanitize@url \@url }%
\providecommand \@url [1]{\endgroup\@href {#1}{\urlprefix }}%
\providecommand \urlprefix  [0]{URL }%
\providecommand \Eprint [0]{\href }%
\providecommand \doibase [0]{http://dx.doi.org/}%
\providecommand \selectlanguage [0]{\@gobble}%
\providecommand \bibinfo  [0]{\@secondoftwo}%
\providecommand \bibfield  [0]{\@secondoftwo}%
\providecommand \translation [1]{[#1]}%
\providecommand \BibitemOpen [0]{}%
\providecommand \bibitemStop [0]{}%
\providecommand \bibitemNoStop [0]{.\EOS\space}%
\providecommand \EOS [0]{\spacefactor3000\relax}%
\providecommand \BibitemShut  [1]{\csname bibitem#1\endcsname}%
\let\auto@bib@innerbib\@empty
%</preamble>
\bibitem [{\citenamefont {Hopfield}(1974)}]{Hopfield1974}%
  \BibitemOpen
  \bibfield  {author} {\bibinfo {author} {\bibfnamefont {J.~J.}\ \bibnamefont
  {Hopfield}},\ }\href {http://www.pnas.org/content/71/10/4135.abstract}
  {\bibfield  {journal} {\bibinfo  {journal} {Proceedings of the National
  Academy of Sciences}\ }\textbf {\bibinfo {volume} {71}},\ \bibinfo {pages}
  {4135} (\bibinfo {year} {1974})}\BibitemShut {NoStop}%
\bibitem [{\citenamefont {Wang}\ \emph {et~al.}(2017)\citenamefont {Wang},
  \citenamefont {Shi}, \citenamefont {He}, \citenamefont {Wang}, \citenamefont
  {Zhang},\ and\ \citenamefont {Yuan}}]{Wang2017}%
  \BibitemOpen
  \bibfield  {author} {\bibinfo {author} {\bibfnamefont {F.}~\bibnamefont
  {Wang}}, \bibinfo {author} {\bibfnamefont {H.}~\bibnamefont {Shi}}, \bibinfo
  {author} {\bibfnamefont {R.}~\bibnamefont {He}}, \bibinfo {author}
  {\bibfnamefont {R.}~\bibnamefont {Wang}}, \bibinfo {author} {\bibfnamefont
  {R.}~\bibnamefont {Zhang}}, \ and\ \bibinfo {author} {\bibfnamefont
  {J.}~\bibnamefont {Yuan}},\ }\href@noop {} {\bibfield  {journal} {\bibinfo
  {journal} {Nature Physics}\ } (\bibinfo {year} {2017})}\BibitemShut {NoStop}%
\bibitem [{\citenamefont {Tu}(2008)}]{Tu2008}%
  \BibitemOpen
  \bibfield  {author} {\bibinfo {author} {\bibfnamefont {Y.}~\bibnamefont
  {Tu}},\ }\href {\doibase 10.1073/pnas.0804641105} {\bibfield  {journal}
  {\bibinfo  {journal} {Proceedings of the National Academy of Sciences of the
  United States of America}\ }\textbf {\bibinfo {volume} {105}},\ \bibinfo
  {pages} {11737} (\bibinfo {year} {2008})}\BibitemShut {NoStop}%
\bibitem [{\citenamefont {Lan}\ \emph {et~al.}(2012)\citenamefont {Lan},
  \citenamefont {Sartori}, \citenamefont {Neumann}, \citenamefont {Sourjik},\
  and\ \citenamefont {Tu}}]{Lan2012}%
  \BibitemOpen
  \bibfield  {author} {\bibinfo {author} {\bibfnamefont {G.}~\bibnamefont
  {Lan}}, \bibinfo {author} {\bibfnamefont {P.}~\bibnamefont {Sartori}},
  \bibinfo {author} {\bibfnamefont {S.}~\bibnamefont {Neumann}}, \bibinfo
  {author} {\bibfnamefont {V.}~\bibnamefont {Sourjik}}, \ and\ \bibinfo
  {author} {\bibfnamefont {Y.}~\bibnamefont {Tu}},\ }\href {\doibase
  10.1038/nphys2276} {\bibfield  {journal} {\bibinfo  {journal} {Nature
  physics}\ }\textbf {\bibinfo {volume} {8}},\ \bibinfo {pages} {422} (\bibinfo
  {year} {2012})}\BibitemShut {NoStop}%
\bibitem [{\citenamefont {Cao}\ \emph {et~al.}(2015)\citenamefont {Cao},
  \citenamefont {Wang}, \citenamefont {Ouyang},\ and\ \citenamefont
  {Tu}}]{Cao2015}%
  \BibitemOpen
  \bibfield  {author} {\bibinfo {author} {\bibfnamefont {Y.}~\bibnamefont
  {Cao}}, \bibinfo {author} {\bibfnamefont {H.}~\bibnamefont {Wang}}, \bibinfo
  {author} {\bibfnamefont {Q.}~\bibnamefont {Ouyang}}, \ and\ \bibinfo {author}
  {\bibfnamefont {Y.}~\bibnamefont {Tu}},\ }\href@noop {} {\bibfield  {journal}
  {\bibinfo  {journal} {Nature physics}\ } (\bibinfo {year}
  {2015})}\BibitemShut {NoStop}%
\bibitem [{\citenamefont {Barato}\ and\ \citenamefont
  {Seifert}(2015)}]{Barato2015}%
  \BibitemOpen
  \bibfield  {author} {\bibinfo {author} {\bibfnamefont {A.~C.}\ \bibnamefont
  {Barato}}\ and\ \bibinfo {author} {\bibfnamefont {U.}~\bibnamefont
  {Seifert}},\ }\href {\doibase 10.1103/PhysRevLett.114.158101} {\bibfield
  {journal} {\bibinfo  {journal} {Physical Review Letters. Rev. Lett.}\
  }\textbf {\bibinfo {volume} {114}},\ \bibinfo {pages} {158101} (\bibinfo
  {year} {2015})}\BibitemShut {NoStop}%
\bibitem [{\citenamefont {Gingrich}\ \emph {et~al.}(2016)\citenamefont
  {Gingrich}, \citenamefont {Horowitz}, \citenamefont {Perunov},\ and\
  \citenamefont {England}}]{Gingrich2016}%
  \BibitemOpen
  \bibfield  {author} {\bibinfo {author} {\bibfnamefont {T.~R.}\ \bibnamefont
  {Gingrich}}, \bibinfo {author} {\bibfnamefont {J.~M.}\ \bibnamefont
  {Horowitz}}, \bibinfo {author} {\bibfnamefont {N.}~\bibnamefont {Perunov}}, \
  and\ \bibinfo {author} {\bibfnamefont {J.~L.}\ \bibnamefont {England}},\
  }\href@noop {} {\bibfield  {journal} {\bibinfo  {journal} {Physical Review
  Letters}\ }\textbf {\bibinfo {volume} {116}},\ \bibinfo {pages} {120601}
  (\bibinfo {year} {2016})}\BibitemShut {NoStop}%
\bibitem [{\citenamefont {Nguyen}\ and\ \citenamefont
  {Vaikuntanathan}(2016)}]{Nguyen2016}%
  \BibitemOpen
  \bibfield  {author} {\bibinfo {author} {\bibfnamefont {M.}~\bibnamefont
  {Nguyen}}\ and\ \bibinfo {author} {\bibfnamefont {S.}~\bibnamefont
  {Vaikuntanathan}},\ }\href@noop {} {\bibfield  {journal} {\bibinfo  {journal}
  {Proceedings of the National Academy of Sciences}\ }\textbf {\bibinfo
  {volume} {113}},\ \bibinfo {pages} {14231} (\bibinfo {year}
  {2016})}\BibitemShut {NoStop}%
\bibitem [{\citenamefont {Kane}\ and\ \citenamefont
  {Lubensky}(2013)}]{Kane2013}%
  \BibitemOpen
  \bibfield  {author} {\bibinfo {author} {\bibfnamefont {C.~L.}\ \bibnamefont
  {Kane}}\ and\ \bibinfo {author} {\bibfnamefont {T.~C.}\ \bibnamefont
  {Lubensky}},\ }\href {\doibase 10.1038/nphys2835} {\bibfield  {journal}
  {\bibinfo  {journal} {Nature Physics}\ }\textbf {\bibinfo {volume} {10}},\
  \bibinfo {pages} {39} (\bibinfo {year} {2013})}\BibitemShut {NoStop}%
\bibitem [{\citenamefont {Hasan}\ and\ \citenamefont {Kane}(2010)}]{Hasan2010}%
  \BibitemOpen
  \bibfield  {author} {\bibinfo {author} {\bibfnamefont {M.~Z.}\ \bibnamefont
  {Hasan}}\ and\ \bibinfo {author} {\bibfnamefont {C.~L.}\ \bibnamefont
  {Kane}},\ }\href {\doibase 10.1103/RevModPhys.82.3045} {\bibfield  {journal}
  {\bibinfo  {journal} {Reviews of Modern Physics}\ }\textbf {\bibinfo {volume}
  {82}},\ \bibinfo {pages} {3045} (\bibinfo {year} {2010})}\BibitemShut
  {NoStop}%
\bibitem [{\citenamefont {Haldane}\ and\ \citenamefont
  {Raghu}(2008)}]{Haldane2008}%
  \BibitemOpen
  \bibfield  {author} {\bibinfo {author} {\bibfnamefont {F.~D.~M.}\
  \bibnamefont {Haldane}}\ and\ \bibinfo {author} {\bibfnamefont
  {S.}~\bibnamefont {Raghu}},\ }\href {\doibase 10.1103/PhysRevLett.100.013904}
  {\bibfield  {journal} {\bibinfo  {journal} {Physical Review Letters}\
  }\textbf {\bibinfo {volume} {100}},\ \bibinfo {pages} {013904} (\bibinfo
  {year} {2008})}\BibitemShut {NoStop}%
\bibitem [{\citenamefont {Nash}\ \emph {et~al.}(2015)\citenamefont {Nash},
  \citenamefont {Kleckner}, \citenamefont {Read}, \citenamefont {Vitelli},
  \citenamefont {Turner},\ and\ \citenamefont {Irvine}}]{Nash2015}%
  \BibitemOpen
  \bibfield  {author} {\bibinfo {author} {\bibfnamefont {L.~M.}\ \bibnamefont
  {Nash}}, \bibinfo {author} {\bibfnamefont {D.}~\bibnamefont {Kleckner}},
  \bibinfo {author} {\bibfnamefont {A.}~\bibnamefont {Read}}, \bibinfo {author}
  {\bibfnamefont {V.}~\bibnamefont {Vitelli}}, \bibinfo {author} {\bibfnamefont
  {A.~M.}\ \bibnamefont {Turner}}, \ and\ \bibinfo {author} {\bibfnamefont
  {W.~T.~M.}\ \bibnamefont {Irvine}},\ }\href {\doibase
  10.1073/pnas.1507413112} {\bibfield  {journal} {\bibinfo  {journal}
  {Proceedings of the National Academy of Sciences}\ }\textbf {\bibinfo
  {volume} {112}},\ \bibinfo {pages} {14495} (\bibinfo {year}
  {2015})}\BibitemShut {NoStop}%
\bibitem [{\citenamefont {Paulose}\ \emph {et~al.}(2015)\citenamefont
  {Paulose}, \citenamefont {Chen},\ and\ \citenamefont
  {Vitelli}}]{Paulose2015}%
  \BibitemOpen
  \bibfield  {author} {\bibinfo {author} {\bibfnamefont {J.}~\bibnamefont
  {Paulose}}, \bibinfo {author} {\bibfnamefont {B.~G.-g.}\ \bibnamefont
  {Chen}}, \ and\ \bibinfo {author} {\bibfnamefont {V.}~\bibnamefont
  {Vitelli}},\ }\href {\doibase 10.1038/nphys3185} {\bibfield  {journal}
  {\bibinfo  {journal} {Nature Physics}\ }\textbf {\bibinfo {volume} {11}},\
  \bibinfo {pages} {153} (\bibinfo {year} {2015})}\BibitemShut {NoStop}%
\bibitem [{\citenamefont {Ryu}\ and\ \citenamefont {Hatsugai}(2002)}]{Ryu2002}%
  \BibitemOpen
  \bibfield  {author} {\bibinfo {author} {\bibfnamefont {S.}~\bibnamefont
  {Ryu}}\ and\ \bibinfo {author} {\bibfnamefont {Y.}~\bibnamefont {Hatsugai}},\
  }\href {\doibase 10.1103/PhysRevLett.89.077002} {\bibfield  {journal}
  {\bibinfo  {journal} {Physical Review Letters}\ }\textbf {\bibinfo {volume}
  {89}},\ \bibinfo {pages} {077002} (\bibinfo {year} {2002})}\BibitemShut
  {NoStop}%
\bibitem [{\citenamefont {Chen}\ \emph {et~al.}(2014)\citenamefont {Chen},
  \citenamefont {Upadhyaya},\ and\ \citenamefont {Vitelli}}]{Chen2014}%
  \BibitemOpen
  \bibfield  {author} {\bibinfo {author} {\bibfnamefont {B.~G.-g.}\
  \bibnamefont {Chen}}, \bibinfo {author} {\bibfnamefont {N.}~\bibnamefont
  {Upadhyaya}}, \ and\ \bibinfo {author} {\bibfnamefont {V.}~\bibnamefont
  {Vitelli}},\ }\href {\doibase 10.1073/pnas.1405969111} {\bibfield  {journal}
  {\bibinfo  {journal} {Proceedings of the National Academy of Sciences}\
  }\textbf {\bibinfo {volume} {111}},\ \bibinfo {pages} {13004} (\bibinfo
  {year} {2014})}\BibitemShut {NoStop}%
\bibitem [{\citenamefont {Lubensky}\ \emph {et~al.}(2015)\citenamefont
  {Lubensky}, \citenamefont {Kane}, \citenamefont {Mao}, \citenamefont
  {Souslov},\ and\ \citenamefont {Sun}}]{Lubensky2015}%
  \BibitemOpen
  \bibfield  {author} {\bibinfo {author} {\bibfnamefont {T.~C.}\ \bibnamefont
  {Lubensky}}, \bibinfo {author} {\bibfnamefont {C.~L.}\ \bibnamefont {Kane}},
  \bibinfo {author} {\bibfnamefont {X.}~\bibnamefont {Mao}}, \bibinfo {author}
  {\bibfnamefont {A.}~\bibnamefont {Souslov}}, \ and\ \bibinfo {author}
  {\bibfnamefont {K.}~\bibnamefont {Sun}},\ }\href
  {http://stacks.iop.org/0034-4885/78/i=7/a=073901} {\bibfield  {journal}
  {\bibinfo  {journal} {Reports on Progress in Physics}\ }\textbf {\bibinfo
  {volume} {78}},\ \bibinfo {pages} {073901} (\bibinfo {year}
  {2015})}\BibitemShut {NoStop}%
\bibitem [{\citenamefont {Murugan}\ \emph {et~al.}(2014)\citenamefont
  {Murugan}, \citenamefont {Huse},\ and\ \citenamefont
  {Leibler}}]{Murugan2014}%
  \BibitemOpen
  \bibfield  {author} {\bibinfo {author} {\bibfnamefont {A.}~\bibnamefont
  {Murugan}}, \bibinfo {author} {\bibfnamefont {D.~A.}\ \bibnamefont {Huse}}, \
  and\ \bibinfo {author} {\bibfnamefont {S.}~\bibnamefont {Leibler}},\ }\href
  {\doibase 10.1103/PhysRevX.4.021016} {\bibfield  {journal} {\bibinfo
  {journal} {Physical Review X.}\ }\textbf {\bibinfo {volume} {4}},\ \bibinfo {pages}
  {021016} (\bibinfo {year} {2014})}\BibitemShut {NoStop}%
\bibitem [{\citenamefont {Tsai}\ \emph {et~al.}(2005)\citenamefont {Tsai},
  \citenamefont {Ye}, \citenamefont {Rodriguez}, \citenamefont {Gollub},\ and\
  \citenamefont {Lubensky}}]{Tsai2005}%
  \BibitemOpen
  \bibfield  {author} {\bibinfo {author} {\bibfnamefont {J.-C.}\ \bibnamefont
  {Tsai}}, \bibinfo {author} {\bibfnamefont {F.}~\bibnamefont {Ye}}, \bibinfo
  {author} {\bibfnamefont {J.}~\bibnamefont {Rodriguez}}, \bibinfo {author}
  {\bibfnamefont {J.~P.}\ \bibnamefont {Gollub}}, \ and\ \bibinfo {author}
  {\bibfnamefont {T.~C.}\ \bibnamefont {Lubensky}},\ }\href {\doibase
  10.1103/PhysRevLett.94.214301} {\bibfield  {journal} {\bibinfo  {journal}
  {Physical Review Letters}\ }\textbf {\bibinfo {volume} {94}},\ \bibinfo
  {pages} {214301} (\bibinfo {year} {2005})}\BibitemShut {NoStop}%
\bibitem [{\citenamefont {van Zuiden}\ \emph {et~al.}(2016)\citenamefont {van
  Zuiden}, \citenamefont {Paulose}, \citenamefont {Irvine}, \citenamefont
  {Bartolo},\ and\ \citenamefont {Vitelli}}]{vanZuiden16}%
  \BibitemOpen
  \bibfield  {author} {\bibinfo {author} {\bibfnamefont {B.~C.}\ \bibnamefont
  {van Zuiden}}, \bibinfo {author} {\bibfnamefont {J.}~\bibnamefont {Paulose}},
  \bibinfo {author} {\bibfnamefont {W.~T.~M.}\ \bibnamefont {Irvine}}, \bibinfo
  {author} {\bibfnamefont {D.}~\bibnamefont {Bartolo}}, \ and\ \bibinfo
  {author} {\bibfnamefont {V.}~\bibnamefont {Vitelli}},\ }\href {\doibase
  10.1073/pnas.1609572113} {\bibfield  {journal} {\bibinfo  {journal}
  {Proceedings of the National Academy of Sciences}\ }\textbf {\bibinfo
  {volume} {113}},\ \bibinfo {pages} {12919} (\bibinfo {year}
  {2016})}\BibitemShut {NoStop}%
\bibitem [{\citenamefont {Su}\ \emph {et~al.}(1979)\citenamefont {Su},
  \citenamefont {Schrieffer},\ and\ \citenamefont {Heeger}}]{SSH79}%
  \BibitemOpen
  \bibfield  {author} {\bibinfo {author} {\bibfnamefont {W.~P.}\ \bibnamefont
  {Su}}, \bibinfo {author} {\bibfnamefont {J.~R.}\ \bibnamefont {Schrieffer}},
  \ and\ \bibinfo {author} {\bibfnamefont {A.~J.}\ \bibnamefont {Heeger}},\
  }\href {\doibase 10.1103/PhysRevLett.42.1698} {\bibfield  {journal} {\bibinfo
   {journal} {Physical Review Letters}\ }\textbf {\bibinfo {volume} {42}},\
  \bibinfo {pages} {1698} (\bibinfo {year} {1979})}\BibitemShut {NoStop}%
\bibitem [{\citenamefont {Schnakenberg}(1976)}]{Schnakenberg1976}%
  \BibitemOpen
  \bibfield  {author} {\bibinfo {author} {\bibfnamefont {J.}~\bibnamefont
  {Schnakenberg}},\ }\href {http://rmp.aps.org/abstract/RMP/v48/i4/p571\_1}
  {\bibfield  {journal} {\bibinfo  {journal} {Reviews of Modern physics}\ }
  (\bibinfo {year} {1976})}\BibitemShut {NoStop}%
\bibitem [{\citenamefont {Van~Kampen}(2011)}]{van2011stochastic}%
  \BibitemOpen
  \bibfield  {author} {\bibinfo {author} {\bibfnamefont {N.}~\bibnamefont
  {Van~Kampen}},\ }\href {https://books.google.com/books?id=N6II-6HlPxEC}
  {\emph {\bibinfo {title} {Stochastic Processes in Physics and Chemistry}}},\
  North-Holland Personal Library\ (\bibinfo  {publisher} {Elsevier Science},\
  \bibinfo {year} {2011})\BibitemShut {NoStop}%
\bibitem [{\citenamefont {Murugan}\ and\ \citenamefont
  {Vaikuntanathan}(2017)}]{Vaikunt2017}%
  \BibitemOpen
  \bibfield  {author} {\bibinfo {author} {\bibfnamefont {A.}~\bibnamefont
  {Murugan}}\ and\ \bibinfo {author} {\bibfnamefont {S.}~\bibnamefont
  {Vaikuntanathan}},\ }\href {\doibase http://dx.doi.org/10.1038/ncomms13881}
  {\bibfield  {journal} {\bibinfo  {journal} {Nature Communications}\ }\textbf
  {\bibinfo {volume} {8}},\ \bibinfo {pages} {13881} (\bibinfo {year}
  {2017})}\BibitemShut {NoStop}%
\bibitem [{\citenamefont {Süsstrunk}\ and\ \citenamefont
  {Huber}(2016)}]{Huber2016}%
  \BibitemOpen
  \bibfield  {author} {\bibinfo {author} {\bibfnamefont {R.}~\bibnamefont
  {Süsstrunk}}\ and\ \bibinfo {author} {\bibfnamefont {S.~D.}\ \bibnamefont
  {Huber}},\ }\href {\doibase 10.1073/pnas.1605462113} {\bibfield  {journal}
  {\bibinfo  {journal} {Proceedings of the National Academy of Sciences}\
  }\textbf {\bibinfo {volume} {113}},\ \bibinfo {pages} {E4767} (\bibinfo
  {year} {2016})}\BibitemShut {NoStop}%
\bibitem [{\citenamefont {Jack}\ and\ \citenamefont
  {Sollich}(2009)}]{Jack2009}%
  \BibitemOpen
  \bibfield  {author} {\bibinfo {author} {\bibfnamefont {R.~L.}\ \bibnamefont
  {Jack}}\ and\ \bibinfo {author} {\bibfnamefont {P.}~\bibnamefont {Sollich}},\
  }\href {http://stacks.iop.org/1742-5468/2009/i=11/a=P11011} {\bibfield
  {journal} {\bibinfo  {journal} {Journal of Statistical Mechanics: Theory and
  Experiment}\ }\textbf {\bibinfo {volume} {2009}},\ \bibinfo {pages} {P11011}
  (\bibinfo {year} {2009})}\BibitemShut {NoStop}%
\bibitem [{\citenamefont {Murugan}\ \emph {et~al.}(2012)\citenamefont
  {Murugan}, \citenamefont {Huse},\ and\ \citenamefont
  {Leibler}}]{Murugan2012}%
  \BibitemOpen
  \bibfield  {author} {\bibinfo {author} {\bibfnamefont {A.}~\bibnamefont
  {Murugan}}, \bibinfo {author} {\bibfnamefont {D.~A.}\ \bibnamefont {Huse}}, \
  and\ \bibinfo {author} {\bibfnamefont {S.}~\bibnamefont {Leibler}},\
  }\href@noop {} {\bibfield  {journal} {\bibinfo  {journal} {Proceedings of the
  National Academy of Sciences}\ }\textbf {\bibinfo {volume} {109}},\ \bibinfo
  {pages} {12034} (\bibinfo {year} {2012})}\BibitemShut {NoStop}%
\bibitem [{\citenamefont {Murugan}\ and\ \citenamefont
  {Vaikuntanathan}(2016)}]{Murugan2016}%
  \BibitemOpen
  \bibfield  {author} {\bibinfo {author} {\bibfnamefont {A.}~\bibnamefont
  {Murugan}}\ and\ \bibinfo {author} {\bibfnamefont {S.}~\bibnamefont
  {Vaikuntanathan}},\ }\href {\doibase 10.1007/s10955-015-1445-0} {\bibfield
  {journal} {\bibinfo  {journal} {Journal of Statistical Physics}\ }\textbf
  {\bibinfo {volume} {162}},\ \bibinfo {pages} {1183} (\bibinfo {year}
  {2016})}\BibitemShut {NoStop}%
\bibitem [{\citenamefont {Mehta}\ and\ \citenamefont
  {Schwab}(2012)}]{Mehta2012}%
  \BibitemOpen
  \bibfield  {author} {\bibinfo {author} {\bibfnamefont {P.}~\bibnamefont
  {Mehta}}\ and\ \bibinfo {author} {\bibfnamefont {D.~J.}\ \bibnamefont
  {Schwab}},\ }\href {\doibase 10.1073/pnas.1207814109} {\bibfield  {journal}
  {\bibinfo  {journal} {Proceedings of the National Academy of Sciences}\
  }\textbf {\bibinfo {volume} {109}},\ \bibinfo {pages} {17978} (\bibinfo
  {year} {2012})}\BibitemShut {NoStop}%
\bibitem [{\citenamefont {Cates}\ and\ \citenamefont
  {Tailleur}(2015)}]{Cates2015}%
  \BibitemOpen
  \bibfield  {author} {\bibinfo {author} {\bibfnamefont {M.~E.}\ \bibnamefont
  {Cates}}\ and\ \bibinfo {author} {\bibfnamefont {J.}~\bibnamefont
  {Tailleur}},\ }\href {\doibase 10.1146/annurev-conmatphys-031214-014710}
  {\bibfield  {journal} {\bibinfo  {journal} {Annual Review of Condensed Matter
  Physics}\ }\textbf {\bibinfo {volume} {6}},\ \bibinfo {pages} {219} (\bibinfo
  {year} {2015})}\BibitemShut {NoStop}%
\bibitem [{\citenamefont {Marchetti}\ \emph {et~al.}(2013)\citenamefont
  {Marchetti}, \citenamefont {Joanny}, \citenamefont {Ramaswamy}, \citenamefont
  {Liverpool}, \citenamefont {Prost}, \citenamefont {Rao},\ and\ \citenamefont
  {Simha}}]{RMP13}%
  \BibitemOpen
  \bibfield  {author} {\bibinfo {author} {\bibfnamefont {M.~C.}\ \bibnamefont
  {Marchetti}}, \bibinfo {author} {\bibfnamefont {J.~F.}\ \bibnamefont
  {Joanny}}, \bibinfo {author} {\bibfnamefont {S.}~\bibnamefont {Ramaswamy}},
  \bibinfo {author} {\bibfnamefont {T.~B.}\ \bibnamefont {Liverpool}}, \bibinfo
  {author} {\bibfnamefont {J.}~\bibnamefont {Prost}}, \bibinfo {author}
  {\bibfnamefont {M.}~\bibnamefont {Rao}}, \ and\ \bibinfo {author}
  {\bibfnamefont {R.~A.}\ \bibnamefont {Simha}},\ }\href {\doibase
  10.1103/RevModPhys.85.1143} {\bibfield  {journal} {\bibinfo  {journal}
  {Reviews of Modern Physics}\ }\textbf {\bibinfo {volume} {85}},\ \bibinfo
  {pages} {1143} (\bibinfo {year} {2013})}\BibitemShut {NoStop}%
\bibitem [{\citenamefont {Kruse}\ \emph {et~al.}(2005)\citenamefont {Kruse},
  \citenamefont {Joanny}, \citenamefont {J{\"u}licher}, \citenamefont {Prost},\
  and\ \citenamefont {Sekimoto}}]{Kruse2005}%
  \BibitemOpen
  \bibfield  {author} {\bibinfo {author} {\bibfnamefont {K.}~\bibnamefont
  {Kruse}}, \bibinfo {author} {\bibfnamefont {J.~F.}\ \bibnamefont {Joanny}},
  \bibinfo {author} {\bibfnamefont {F.}~\bibnamefont {J{\"u}licher}}, \bibinfo
  {author} {\bibfnamefont {J.}~\bibnamefont {Prost}}, \ and\ \bibinfo {author}
  {\bibfnamefont {K.}~\bibnamefont {Sekimoto}},\ }\href {\doibase
  10.1140/epje/e2005-00002-5} {\bibfield  {journal} {\bibinfo  {journal} {The
  European Physical Journal E}\ }\textbf {\bibinfo {volume} {16}},\ \bibinfo
  {pages} {5} (\bibinfo {year} {2005})}\BibitemShut {NoStop}%
\bibitem [{\citenamefont {Whitelam}\ \emph {et~al.}(2012)\citenamefont
  {Whitelam}, \citenamefont {Schulman},\ and\ \citenamefont
  {Hedges}}]{Whitelam2012}%
  \BibitemOpen
  \bibfield  {author} {\bibinfo {author} {\bibfnamefont {S.}~\bibnamefont
  {Whitelam}}, \bibinfo {author} {\bibfnamefont {R.}~\bibnamefont {Schulman}},
  \ and\ \bibinfo {author} {\bibfnamefont {L.}~\bibnamefont {Hedges}},\ }\href
  {\doibase 10.1103/PhysRevLett.109.265506} {\bibfield  {journal} {\bibinfo
  {journal} {Physical Review Letters}\ }\textbf {\bibinfo {volume} {109}},\
  \bibinfo {pages} {265506} (\bibinfo {year} {2012})}\BibitemShut {NoStop}%
\bibitem [{\citenamefont {de~Groot}\ and\ \citenamefont
  {Mazur}(1969)}]{deGroot1969}%
  \BibitemOpen
  \bibfield  {author} {\bibinfo {author} {\bibfnamefont {S.}~\bibnamefont
  {de~Groot}}\ and\ \bibinfo {author} {\bibfnamefont {P.}~\bibnamefont
  {Mazur}},\ }\href {https://books.google.com/books?id=U3DVcQAACAAJ} {\emph
  {\bibinfo {title} {Non-equilibrium Thermodynamics: By S.R. de Groot and P.
  Mazur}}}\ (\bibinfo  {publisher} {North-Holland},\ \bibinfo {year}
  {1969})\BibitemShut {NoStop}%
\bibitem [{\citenamefont {F\"urthauer}\ \emph {et~al.}(2013)\citenamefont
  {F\"urthauer}, \citenamefont {Strempel}, \citenamefont {Grill},\ and\
  \citenamefont {J\"ulicher}}]{Strempel2013}%
  \BibitemOpen
  \bibfield  {author} {\bibinfo {author} {\bibfnamefont {S.}~\bibnamefont
  {F\"urthauer}}, \bibinfo {author} {\bibfnamefont {M.}~\bibnamefont
  {Strempel}}, \bibinfo {author} {\bibfnamefont {S.~W.}\ \bibnamefont {Grill}},
  \ and\ \bibinfo {author} {\bibfnamefont {F.}~\bibnamefont {J\"ulicher}},\
  }\href {\doibase 10.1103/PhysRevLett.110.048103} {\bibfield  {journal}
  {\bibinfo  {journal} {Physical Review Letters}\ }\textbf {\bibinfo {volume}
  {110}},\ \bibinfo {pages} {048103} (\bibinfo {year} {2013})}\BibitemShut
  {NoStop}%
\bibitem [{\citenamefont {Nguyen}\ \emph {et~al.}(2014)\citenamefont {Nguyen},
  \citenamefont {Klotsa}, \citenamefont {Engel},\ and\ \citenamefont
  {Glotzer}}]{Nguyen2014}%
  \BibitemOpen
  \bibfield  {author} {\bibinfo {author} {\bibfnamefont {N.~H.~P.}\
  \bibnamefont {Nguyen}}, \bibinfo {author} {\bibfnamefont {D.}~\bibnamefont
  {Klotsa}}, \bibinfo {author} {\bibfnamefont {M.}~\bibnamefont {Engel}}, \
  and\ \bibinfo {author} {\bibfnamefont {S.~C.}\ \bibnamefont {Glotzer}},\
  }\href {\doibase 10.1103/PhysRevLett.112.075701} {\bibfield  {journal}
  {\bibinfo  {journal} {Physical Review Letters}\ }\textbf {\bibinfo {volume}
  {112}},\ \bibinfo {pages} {075701} (\bibinfo {year} {2014})}\BibitemShut
  {NoStop}%
\bibitem [{\citenamefont {Migler}\ and\ \citenamefont
  {Meyer}(1991)}]{Migler1991}%
  \BibitemOpen
  \bibfield  {author} {\bibinfo {author} {\bibfnamefont {K.~B.}\ \bibnamefont
  {Migler}}\ and\ \bibinfo {author} {\bibfnamefont {R.~B.}\ \bibnamefont
  {Meyer}},\ }\href {\doibase 10.1103/PhysRevLett.66.1485} {\bibfield
  {journal} {\bibinfo  {journal} {Physical Review Letters}\ }\textbf {\bibinfo
  {volume} {66}},\ \bibinfo {pages} {1485} (\bibinfo {year}
  {1991})}\BibitemShut {NoStop}%
\bibitem [{\citenamefont {Maggi}\ \emph {et~al.}(2015)\citenamefont {Maggi},
  \citenamefont {Saglimbeni}, \citenamefont {Dipalo}, \citenamefont
  {De~Angelis},\ and\ \citenamefont {Di~Leonardo}}]{Maggi2015}%
  \BibitemOpen
  \bibfield  {author} {\bibinfo {author} {\bibfnamefont {C.}~\bibnamefont
  {Maggi}}, \bibinfo {author} {\bibfnamefont {F.}~\bibnamefont {Saglimbeni}},
  \bibinfo {author} {\bibfnamefont {M.}~\bibnamefont {Dipalo}}, \bibinfo
  {author} {\bibfnamefont {F.}~\bibnamefont {De~Angelis}}, \ and\ \bibinfo
  {author} {\bibfnamefont {R.}~\bibnamefont {Di~Leonardo}},\ }\href@noop {}
  {\bibfield  {journal} {\bibinfo  {journal} {Nature Communications}\ }\textbf
  {\bibinfo {volume} {6}},\ \bibinfo {pages} {7855} (\bibinfo {year}
  {2015})}\BibitemShut {NoStop}%
\bibitem [{\citenamefont {Naganathan}\ \emph {et~al.}(2014)\citenamefont
  {Naganathan}, \citenamefont {F{\"u}rthauer}, \citenamefont {Nishikawa},
  \citenamefont {J{\"u}licher},\ and\ \citenamefont {Grill}}]{Naganathan2014}%
  \BibitemOpen
  \bibfield  {author} {\bibinfo {author} {\bibfnamefont {S.~R.}\ \bibnamefont
  {Naganathan}}, \bibinfo {author} {\bibfnamefont {S.}~\bibnamefont
  {F{\"u}rthauer}}, \bibinfo {author} {\bibfnamefont {M.}~\bibnamefont
  {Nishikawa}}, \bibinfo {author} {\bibfnamefont {F.}~\bibnamefont
  {J{\"u}licher}}, \ and\ \bibinfo {author} {\bibfnamefont {S.~W.}\
  \bibnamefont {Grill}},\ }\href {\doibase 10.7554/eLife.04165} {\bibfield
  {journal} {\bibinfo  {journal} {eLife}\ }\textbf {\bibinfo {volume} {3}},\
  \bibinfo {pages} {e04165} (\bibinfo {year} {2014})}\BibitemShut {NoStop}%
\bibitem [{\citenamefont {F{\"u}rthauer}\ \emph {et~al.}(2012)\citenamefont
  {F{\"u}rthauer}, \citenamefont {Strempel}, \citenamefont {Grill},\ and\
  \citenamefont {J{\"u}licher}}]{Furthauer2012}%
  \BibitemOpen
  \bibfield  {author} {\bibinfo {author} {\bibfnamefont {S.}~\bibnamefont
  {F{\"u}rthauer}}, \bibinfo {author} {\bibfnamefont {M.}~\bibnamefont
  {Strempel}}, \bibinfo {author} {\bibfnamefont {S.~W.}\ \bibnamefont {Grill}},
  \ and\ \bibinfo {author} {\bibfnamefont {F.}~\bibnamefont {J{\"u}licher}},\
  }\href {\doibase 10.1140/epje/i2012-12089-6} {\bibfield  {journal} {\bibinfo
  {journal} {The European Physical Journal E}\ }\textbf {\bibinfo {volume}
  {35}},\ \bibinfo {pages} {89} (\bibinfo {year} {2012})}\BibitemShut {NoStop}%
\bibitem [{\citenamefont {Klymko}\ \emph
  {et~al.}(2017{\natexlab{a}})\citenamefont {Klymko}, \citenamefont {Mandal},\
  and\ \citenamefont {Mandadapu}}]{Mandal2017}%
  \BibitemOpen
  \bibfield  {author} {\bibinfo {author} {\bibfnamefont {K.}~\bibnamefont
  {Klymko}}, \bibinfo {author} {\bibfnamefont {D.}~\bibnamefont {Mandal}}, \
  and\ \bibinfo {author} {\bibfnamefont {K. K.}~\bibnamefont {Mandadapu}},\
  }\href@noop {} {\bibfield  {journal} {\bibinfo  {journal} {arXiv preprint
  arXiv:1706.02284v1}\ } (\bibinfo {year} {2017}{\natexlab{a}})}\BibitemShut
  {NoStop}%
\bibitem [{\citenamefont {Klymko}\ \emph
  {et~al.}(2017{\natexlab{b}})\citenamefont {Klymko}, \citenamefont {Mandal},\
  and\ \citenamefont {Mandadapu}}]{Klymko2017}%
  \BibitemOpen
  \bibfield  {author} {\bibinfo {author} {\bibfnamefont {K.}~\bibnamefont
  {Klymko}}, \bibinfo {author} {\bibfnamefont {D.}~\bibnamefont {Mandal}}, \
  and\ \bibinfo {author} {\bibfnamefont {K. K.}~\bibnamefont {Mandadapu}},\
  }\href@noop {} {\bibfield  {journal} {\bibinfo  {journal} {arXiv:1706.02694}\
  } (\bibinfo {year} {2017}{\natexlab{b}})}\BibitemShut {NoStop}%
\bibitem [{\citenamefont {Rocklin}\ \emph {et~al.}(2016)\citenamefont
  {Rocklin}, \citenamefont {Chen}, \citenamefont {Falk}, \citenamefont
  {Vitelli},\ and\ \citenamefont {Lubensky}}]{Lubensky2016}%
  \BibitemOpen
  \bibfield  {author} {\bibinfo {author} {\bibfnamefont {D.~Z.}\ \bibnamefont
  {Rocklin}}, \bibinfo {author} {\bibfnamefont {B.~G.}\ \bibnamefont {Chen}},
  \bibinfo {author} {\bibfnamefont {M.}~\bibnamefont {Falk}}, \bibinfo {author}
  {\bibfnamefont {V.}~\bibnamefont {Vitelli}}, \ and\ \bibinfo {author}
  {\bibfnamefont {T.~C.}\ \bibnamefont {Lubensky}},\ }\href {\doibase
  10.1103/PhysRevLett.116.135503} {\bibfield  {journal} {\bibinfo  {journal}
  {Physical Review Letters}\ }\textbf {\bibinfo {volume} {116}},\ \bibinfo
  {pages} {135503} (\bibinfo {year} {2016})}\BibitemShut {NoStop}%
\bibitem [{\citenamefont {Souslov}\ \emph {et~al.}(2016)\citenamefont
  {Souslov}, \citenamefont {van Zuiden}, \citenamefont {Bartolo},\ and\
  \citenamefont {Vitelli}}]{Souslov2016}%
  \BibitemOpen
  \bibfield  {author} {\bibinfo {author} {\bibfnamefont {A.}~\bibnamefont
  {Souslov}}, \bibinfo {author} {\bibfnamefont {B.~C.}\ \bibnamefont {van
  Zuiden}}, \bibinfo {author} {\bibfnamefont {D.}~\bibnamefont {Bartolo}}, \
  and\ \bibinfo {author} {\bibfnamefont {V.}~\bibnamefont {Vitelli}},\
  }\href@noop {} {\bibfield  {journal} {\bibinfo  {journal} {arXiv preprint
  arXiv:1610.06873}\ } (\bibinfo {year} {2016})}\BibitemShut {NoStop}%
\bibitem [{\citenamefont {Shankar}\ \emph {et~al.}(2017)\citenamefont
  {Shankar}, \citenamefont {Bowick},\ and\ \citenamefont
  {Marchetti}}]{Shankar2017}%
  \BibitemOpen
  \bibfield  {author} {\bibinfo {author} {\bibfnamefont {S.}~\bibnamefont
  {Shankar}}, \bibinfo {author} {\bibfnamefont {M.~J.}\ \bibnamefont {Bowick}},
  \ and\ \bibinfo {author} {\bibfnamefont {M.~C.}\ \bibnamefont {Marchetti}},\
  }\href@noop {} {\bibfield  {journal} {\bibinfo  {journal} {arXiv preprint
  arXiv:1704.05424}\ } (\bibinfo {year} {2017})}\BibitemShut {NoStop}%
\end{thebibliography}
%merlin.mbs apsrev4-1.bst 2010-07-25 4.21a (PWD, AO, DPC) hacked
%Control: key (0)
%Control: author (72) initials jnrlst
%Control: editor formatted (1) identically to author
%Control: production of article title (-1) disabled
%Control: page (0) single
%Control: year (1) truncated
%Control: production of eprint (0) enabled
%

\end{document}

\title{Topological localization in  out-of-equilibrium dissipative systems: Supporting Information}
\author{Kinjal Dasbiswas$^1$, Kranthi K. Mandadapu$^{2,3}$, Suriyanarayanan Vaikuntanathan$^{1,4}$} 
\affiliation{$^1$The James Franck Institute, The University of Chicago, Chicago, IL,}
\affiliation{$^2$ Department of Chemical and Biomolecular Engineering, University of California, Berkeley, Berkeley, CA,}
\affiliation{$^3$ Chemical  Sciences Division, Lawrence Berkeley National Laboratory, Berkeley, CA,}
\affiliation{$^4$ Department of Chemistry, The University of Chicago, Chicago, IL.}
\maketitle 

\setcounter{equation}{0}
\setcounter{figure}{0}
\setcounter{table}{0}
\setcounter{page}{1}
\setcounter{section}{0}
\setcounter{subsection}{0}
\makeatletter
\renewcommand{\theequation}{S\arabic{equation}}
\renewcommand{\thefigure}{S\arabic{figure}}
\renewcommand{\bibnumfmt}[1]{[S#1]}
\renewcommand{\citenumfont}[1]{S#1}

\begin{figure}[tbp]
	%\begin{center}
	\includegraphics[width = 5 in]{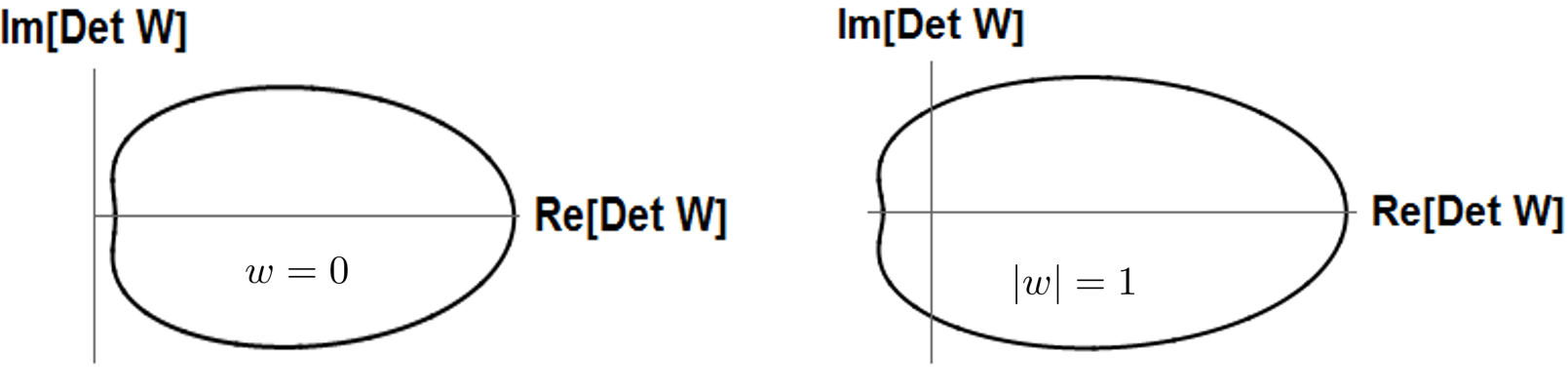}
	\caption{ {\bf Winding number of effective rate matrix} Plots of the determinant of $\tilde{W}_{1}^{x} (k, \lambda)$  as $k$ is varied from $0$ to $2 \pi$ showing two different winding numbers. The plot on the left is for $u/d = 5$ whereas the right is for $u/d = 1/5$. The other rates are chosen to generate a polarization in the probability flow: $ l^{U}/r^{U} = r^{L}/l^{L} = 2$, and small but finite bias: $\lambda \ll 1$. }
	\label{fig:S1}
\end{figure}

\begin{figure}[tbp]
	%\begin{center}
	\includegraphics[width = 4 in]{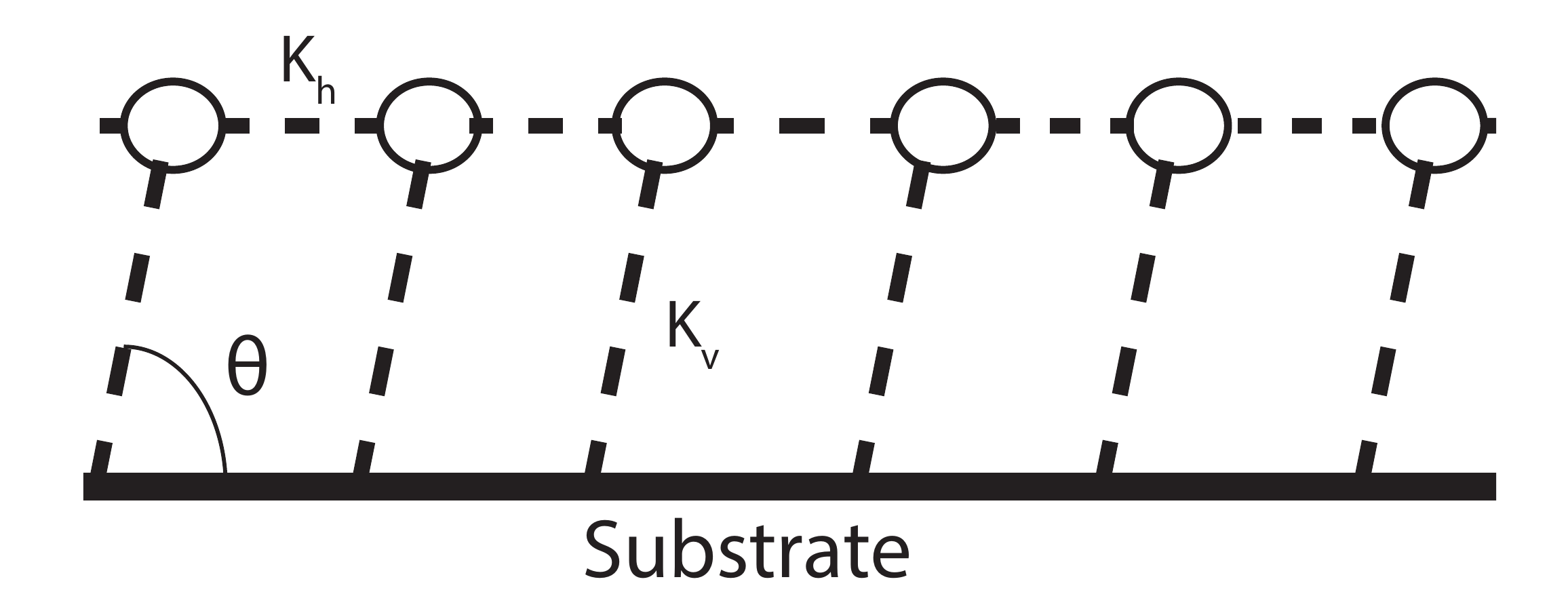}
	\caption{An elastic spring network with modes that resemble those of the linear hydrodynamic operator described in Eq 13 of the main text. By Considering fluctuations about the angle $\theta$, it can be shown the the system has non-trivial topological modes for $\theta\neq \pi/2 $. The springs connecting the chain to the substrate exert a horizontal force on average and this force is responsible for the effective polarization in this model. The action of these 
		springs mimics the action of friction in our hydrodynamic model.}
	\label{fig:spring}
\end{figure}

\begin{figure}[h]
	%\begin{center}
	\includegraphics[width = 6 in]{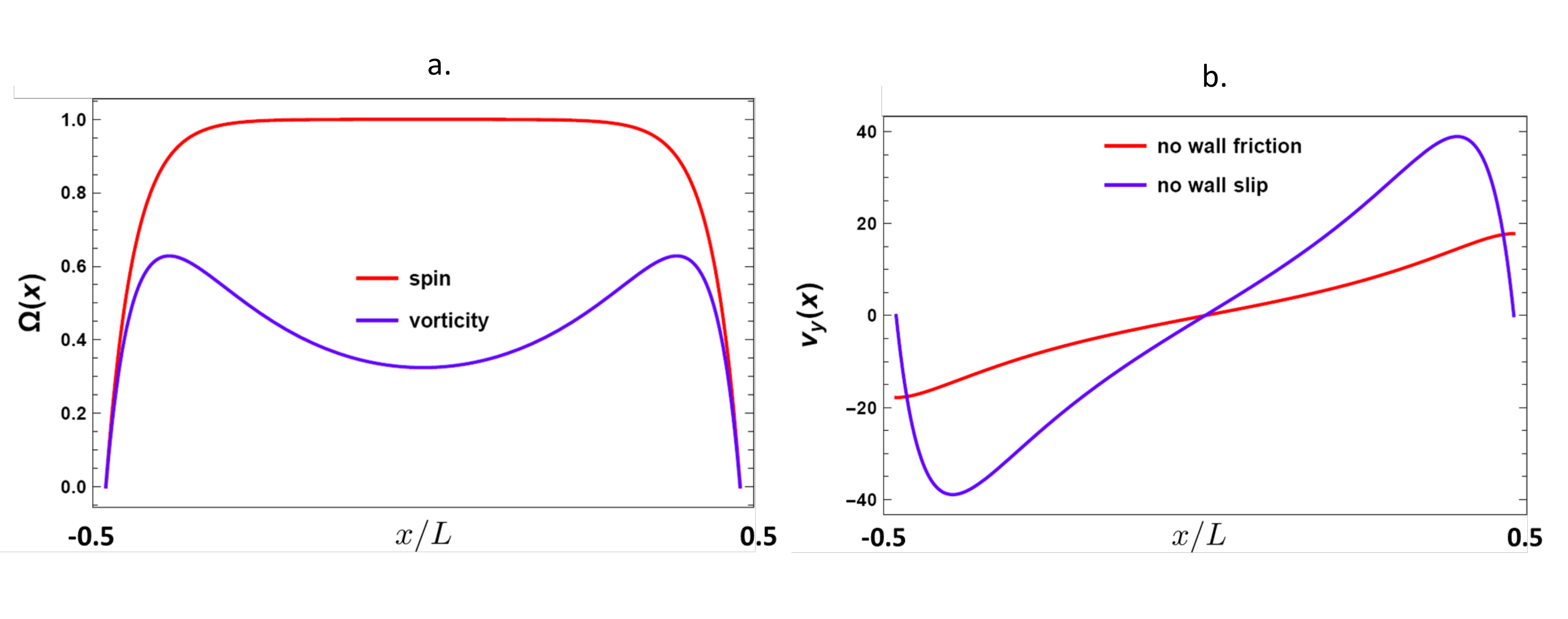}
	\caption{ {\bf Hydrodynamic  profiles} Plots of profiles of hydrodynamic variables of the fluid of spinners as a function of distance from the walls in a slab geometry (section taken along $x-$ direction). a. Comparison of fluid spin, $\Omega(x)$, and vorticity, $\omega(x)$. The bulk spin value, $\Omega_{\rm{b}}$ is normalized. The parameter values used are: $\lambda_{\Omega} = 3.5$, $\lambda_{\omega} = 22.0$, and box size, $L=80$. 
	b. the ($y$-component of) velocity  for both ``no slip'' and ``zero tangential force'' at wall boundary conditions. These profiles are quantitatively different, but both can be interpreted as topological modes localized at the boundaries. }
	\label{fig:S2}
\end{figure}

\section*{Topological modes in master equations}

\subsection*{Supp. Note. 1:  Bulk-boundary correspondence for biased random walk in 1D}

%We consider a master equation, $\dot{{\bf P}} =  {\boldsymbol W}  {\boldsymbol P}$,: \bf allows italic/slanted bold mathematical symbols. 

We consider a master equation, $\dot{{\bf P}} =  {\bf W}  {\bf P}$, where the probability vector, ${\bf P}(t)$, is evolved in time by a transition matrix, ${\bf W}$. The elements of the probability vector, $P_{i}$, denote the probability that the $i^{th}$ node is occupied at a given time. In the convention we adopt here and henceforth, the probability is a column vector, and the rate matrix elements $W_{ij} $ describe transition rates from the $j^{th}$ to $i^{th}$ node.

For a Markovian random process in 1-dimension (1D), as shown in Fig.1 in the main text), the probability and current vectors, ${\bf P}$ and  ${\bf J}$ are related as,
\begin{eqnarray}
\dot{P}_{i} &=& J_{i} - J_{i+1}  \equiv (W_{0})_{i,j} J_{i}, \nonumber \\
J_{i} &=& -v P_{i} + w P_{i-1}  \equiv (W_{1})_{i,j} P_{i},  \nonumber \\
or \, \dot{P}_{i} &=& w P_{i} - (v+w) P_{i} + v P_{i+1}
\label{eq:operator}
\end{eqnarray}
where $v$ and $w$ are the rates of backward and forward  jumps respectively.  This corresponds to a random walk in 1D which is biased by an external potential such that in general, $v \neq w$.
This decomposition can be written explicitly in real space as,
\begin{equation}
{\bf W} = \begin{pmatrix} 
 w & -(v+w) & v & .\\  0 & w & -(v+w) & v\\ . & . & . & .  \end{pmatrix} = {\bf  W_{0}} \cdot {\bf W_{1}}
 = \begin{pmatrix} 
 1 & -1 & 0 & .\\  0 & 1 & -1 & 0\\ . & . & . & .  \end{pmatrix} \cdot
 \begin{pmatrix} -v & 0 & 0 & .\\ w & -v & 0 & .\\ 0 & w & -v & .  \end{pmatrix},
 \label{eq:decomp} 
\end{equation}
where ${\bf W_{0}}$ depends on the connectivity of the network and ${\bf W_{1}}$ contains the hopping probabilities. It is easy to see that the elements of each column add to zero or equivalently that $\bra{1 1 .. 1}$ is a left zero eigenvector of ${\bf W}$. This ensures conservation of probability (Kirchoff's rule). If ${ \bf W_{1}}$ has a zero right eigenvector, this is also a right zero eigenvector of $W$ by construction. According to the Perron-Frobenius theorem \cite{van2011stochastic}, ${\bf W}$ has exactly one zero right eigenvector which determines the steady state of a master equation. Thus if we can show that ${\bf W}$ above has a right zero eigenvector, this has to be the unique steady state probability vector.

In the main text (Eq. 4), we use the ${\bf W_{1}}$ obtained from the decomposition of the master equation, Eq.~\ref{eq:decomp}, to construct a Hermitian matrix, ${\bf H} = \begin{pmatrix} 0 & {\bf W_{1}}\\  {\bf W_{1}}^{T} & 0 \end{pmatrix}$, which is isomorphic to a 1D tight-binding model (SSH) Hamiltonian. 

The zero eigenvectors of ${\bf H}$ include zero eigenvectors of both ${\bf W_{1}}$ and ${\bf W_{1}}^{T}$.  The topological index theorem has been proved for ${\bf H}$ and shows that the difference in number of localized zero eigenvectors of ${\bf W}$ and ${\bf W}^{T}$ at an interface can be related to the difference of bulk winding numbers on either side of the interface, $\delta \nu$ \cite{Kane2013}. The difference in bulk winding numbers for the 1D system is at most $1$. Here we show by explicit construction that the zero mode of ${\bf W_{1}}$ is located at the interface between the bulk regions or at the opposite ends of these regions (Fig. 1a \& 1c in the main text). For this, it is sufficient to show that ${\bf W_{1}}$ and ${\bf W_{1}}^{T}$ have zero modes localized at opposite ends of a 1D Markov chain. In order to show this, note that, 
\begin{equation}
\begin{pmatrix} -v & 0 & . & .\\ . & . & . & .\\ . & w & -v & 0 \\. & 0 & w & -v \end{pmatrix}
\begin{pmatrix} . \\ r^{2} \\r \\ 1 \end{pmatrix} = 0,
\end{equation}
is satisfied by construction if $r=v/w < 1$  and $N$ is large. In other words, we have identified a zero mode of ${\bf W_{1}}$ localized at one end of a finite domain. This zero mode can be written as $\ket{r^{N}...r^{2} \, r \, 1}$.  Now, consider the corresponding transpose, ${\bf W_{1}}^{T}$. It is easy to see that,
\begin{equation}
\begin{pmatrix} -v & w & 0 & .\\  0 & -v & w & .\\ . & . & . & . \\. & . & 0 & -v \end{pmatrix}
\begin{pmatrix} 1 \\ r \\r^{2} \\ . \end{pmatrix} = 0,
\end{equation}
\emph{i.e.} $\ket{1 \, r \, r^{2} . . r^{N}}$ is a zero mode of ${\bf W_{1}}^{T}$.  Thus it is easy to see that the zero modes of ${\bf W_{1}}$ and  $W_{1}^{T}$ are localized at opposite ends of a 1D Markov chain.  This is analogous to the 1D SSH model for both electronic and mechanical systems\cite{Kane2013}.  In the latter, ${\bf Q}$ and ${\bf Q}^{T}$, the equilibrium and compatibility matrices, play the roles of ${\bf W_{1}}$ and  ${\bf W_{1}}^{T}$, and the zero modes localized at opposite ends correspond to states of self stress and zero energy states respectively \cite{Kane2013}. In the quantum SSH model for electrons, this is analogous to localization on one of two sublattices (shown in red and white in Fig. 1).

\subsection*{Supp. Note 2: Decomposition of master equation for ladder network}

In this note, we  establish the details of topological localization for the case of ladder network shown in Fig. 2a of the main text.  We construct a $2N \times 1$ column vector to represent the ladder where the first $N$ entries represent the upper rail and the next $N$ correspond to the lower rail (See Fig. 2a). The master equation can be written as,
\begin{eqnarray}
\dot{P}^{U}_{i} &\equiv& \dot{P}_{i} = J^{U}_{i} - J^{U}_{i+1} + J_{y} \nonumber \\
\dot{P}^{L}_{i} &\equiv& \dot{P}_{i+N} = J^{L}_{i} - J^{L}_{i+1} - J_{y} 
\end{eqnarray}
where the superscripts, $U$ and $L$, represent the upper and lower rails respectively, and the horizontal current along a rail is (same as in the 1D random walk considered before): $ J^{U}_{i} = -l^{U} P^{U}_{i} + r^{U} P^{U}_{i+1} $, $ J^{L}_{i} = -l^{L} P^{L}_{i} + r^{L} P^{L}_{i+1} $, while
$J_{y} = u P^{L}_{i} - d P^{U}_{i}$ is the net upward vertical current.  Along each  rail, \emph{i.e.} in the $x-$direction, the master equation can be factorized into a connectivity-dependent, ${\bf W}_{0}^{x}$, and a rate-dependent, ${\bf W}_{1}^{x}$, part as in the 1D case above, while there is a contribution to the transition matrix that couples the upper and lower rails.  The transition matrix can therefore be decomposed as,
\begin{equation}
{\bf W} = {\bf W}_{0}^{x} \cdot {\bf W}_{1}^{x} + {\bf W}^{y}
 =\begin{pmatrix} 
 1 & -1 & 0 & .\\  0 & 1 & -1 & 0\\  0 & 0 & 1 & -1  \\ . & . & . & .  \end{pmatrix} \cdot
 \begin{pmatrix} -l^{U} & 0 & 0 & .\\ r^{U} & -l^{U} & 0 & .\\ 0 & -l^{L} & -l^{L} & 0\\0 &  0 & r^{L} & -l^{L}  \end{pmatrix} + 
 \begin{pmatrix} -d & 0 & u & .\\ 0 & -d & 0 & u\\ d & 0 & -u & 0\\0 & d  & 0 & -u \end{pmatrix}.
 \label{eq:decomp2} 
\end{equation} 
Since the bare ${\bf W}_{0}$ matrix is singular,  we regularize it by biasing the rates of the horizontal flow with a factor of $e^{\lambda}$ in the forward (right) direction, such that the inverse matrix, ${\bf W}_{0}^{x}(\lambda))^{-1}$ is well-defined for $\lambda > 0$. We can then factor out the ${\bf W}_{0}^{x}$ matrix which contains no information about the transition rates and decompose the transition matrix as, ${\bf W} \equiv {\bf W}_{0}^{x} \cdot \tilde{{\bf W}}_{1}^{x}$, where the matrix that effectively contains the rate constants is given by, $\tilde{{\bf W}}_{1}^{x} \equiv {\bf W}_{1}^{x} + ({\bf W}_{0}^{x})^{-1} {\bf W}^{y}$.  These $2N \times 2N$ matrices consist of four $N \times N$ blocks with constant elements, \emph{i.e.} they correspond to two coupled sublattices with periodic order. They can therefore be diagonalized in a Fourier basis into $2 \times 2$ matrices as,
\begin{equation}
{\bf W}(\lambda, k) = 
\begin{pmatrix} e^{\lambda} -e^{-\lambda + i k} & 0 \\ 0 & e^{\lambda} - e^{-\lambda + i k}\end{pmatrix} \cdot
\begin{pmatrix} -l^{U} + r^{U} e^{-i k} & 0 \\ 0 & -l^{L} + r^{L} e^{-i k}\end{pmatrix} +
\begin{pmatrix} -d & u \\ d & -u \end{pmatrix}.
\label{eq:}
\end{equation}
Factoring out $W_{0}^{x}$, the resulting effective matrix in Fourier space is then,

\begin{equation}
\tilde{{\bf W}}_{1}^{x} (\lambda,k)= {\bf W}_{1}^{x} + ({\bf W}_{0}^{x})^{-1} {\bf W}^{y} =
\begin{pmatrix} 
-l^{U}  + r^{U} e^{- i k} - d \cdot f^{-1}(k,\lambda) & u \cdot f^{-1}(k,\lambda)\\
d\cdot f^{-1}(k,\lambda) & -l^{L} + r^{L} e^{ -i k} - u \cdot f^{-1}(k,\lambda)
\end{pmatrix},
\label{def:W1eff}
\end{equation}
where $f(k, \lambda)= e^{\lambda} - e^{-\lambda + i k}$. 
%is a factor corresponding to the Fourier transform of $W^{x}_{0}$.

\subsection*{ Supp. Note. 3: Topological properties of the effective rate matrix}

We now illustrate that $\tilde{{\bf W}}_{1}^{x} (k, \lambda)$ above encodes topological information in the form of bulk winding numbers.  Analogous to the 1D Markov chain in the previous section, we construct a Hermitian operator of the form,
\begin{equation}
\tilde{\bf H} =
\begin{pmatrix} 
 0 &\tilde{{\bf W}}_{1}^{x}\\
 (\tilde{{\bf W}}_{1}^{x})^{\rm T} & 0
\end{pmatrix},
%= \begin{pmatrix} 
%0 & -v+ w e^{i k} \\  -v + w e^{-i k}  & 0 
%\end{pmatrix} 
\label{defineH2}
\end{equation}
which is set up to have the same zero eigenvectors as $\tilde{{\bf W}}_{1}^{x}$.  Again the trivial zero eigenvector of the original master equation, ${\bf W}$, is not an eigenvector of $\tilde{{\bf W}}_{1}^{x}$ in general. Therefore, any topologically guaranteed zero modes of $\tilde{{\bf H}}$ are also guaranteed to be zero modes of $\tilde{{\bf W}}_{1}^{x}$ because of an inherent bulk-edge correspondence \cite{Kane2013}.

The Hermitian operator defined in Eq.~\ref{defineH2} describes the Hamiltonian for a tight-binding model for electrons on a lattice.  Just as a 1D Markov chain can be mapped to a 1D SSH model for electrons on a lattice with identical topologically protected localized zero eigenmodes, the ladder Markov network maps to two coupled SSH chains with higher order couplings between next nearest neighbors. These nonlocal hopping rates are exponentially small in the bias, being of the order of $\exp(-2 n \lambda)$ for the $n^{th}$ neighbor.  For higher values of the bias, $\lambda \gg 1$, the higher order couplings can be neglected and the analogous tight-binding Hamiltonian assumes a simple form.  This can be seen by expanding the matrix elements in Eq.~(\ref{def:W1eff}) in small $e^{-\lambda}$:
\begin{eqnarray}
\tilde{\bf W}_{1}^{x} (\lambda,k) &=&
\begin{pmatrix} 
-l^{U}  + r^{U} e^{- i k} - d e^{-\lambda} \Sigma_{n} e^{i k n} e^{-2 n \lambda} & u \cdot e^{-\lambda} \Sigma_{n} e^{i k n} e^{-2 n \lambda}\\
d\cdot e^{-\lambda} \Sigma_{n} e^{i k n} e^{-2 n \lambda} & -l^{L} + r^{L} e^{ -i k} - u \cdot e^{-\lambda} \Sigma_{n} e^{i k n} e^{-2 n \lambda}
\end{pmatrix} \nonumber
\\
&\simeq& 
\begin{pmatrix} 
-l^{U}  + r^{U} e^{- i k}  & u e^{-\lambda} \\
d e^{-\lambda} & -l^{L} + r^{L} e^{ -i k} - u \cdot e^{-\lambda} 
\end{pmatrix} \forall \, \lambda \gg 1.
\label{exp:W1eff}
\end{eqnarray}
This tight-binding model corresponding to the matrix in Eq.~\ref{exp:W1eff} schematically represented in Fig. 2b in the main text. Note that the transverse couplings between the two SSH chains is exponentially small in $\lambda$. This is a reason why for large $\lambda$, the network polarization and therefore, topological protection, is lost.

We choose rate constants that provide a net polarization to the direction of probability flow in the ladder network (Fig. 2a in the main text): $l^{U}/r^{U} < 1$ and $l^{L}/r^{L} > 1$. We see that the winding number of the effective rate constant matrix, $\tilde{{\bf W}}_{1}^{x} (k, \lambda)$, defined in terms of the number of times the phase of its determinant winds around the origin in complex space as $k$ is varied from $0$ to $2 \pi$, can be either $0$ or $1$ depending on the ratio $u/d$ of the transition rates in the vertical $y-$direction.

\section*{Topological modes in hydrodynamics}

\subsection*{ Supp. Note 3: Bulk-boundary correspondence for simple fluid model}

First we show that the simple operator, $ \nabla^{2} - \lambda^{-2} $, which occurs generically in dissipative hydrodynamics and in the illustrative example in the main text, can be characterized in terms of a topological index which predicts the number of localized steady state solutions.  For this, we use the identity that any positive symmetric operator can be decomposed in the form, ${\bf D} {\bf D}^{\rm T}$.  This lets us define a ``square-root'' matrix: $
{\bf S} = \begin{pmatrix}  0 & {\bf D} \\ {\bf D} ^{T} & 0 \end{pmatrix}$ which has ``particle-hole'' symmetry, $ \{\sigma_{z},S\} =0$, where $\sigma_{z}$ is a block Pauli matrix \cite{Kane2013}. Eigenmodes (except zero eigenmodes) therefore occur in pairs of positive and negative eigenvalues.
%The matrix $S^{2}$ includes the zero modes of both $D_{0}$ and $D_{0}^{T}$

We now show explicitly how such a square-root decomposition, $ {\bf D} {\bf D}^{T}$, may be achieved for a finite ($N \times N$) realization of the dynamical operator of interest. For this, we write the discretized form of the Laplacian operator using a 1D mesh of size, $h$,
\begin{eqnarray}
 -(\nabla^{2} - \lambda^{-2} ) &=& -\begin{pmatrix} 
 -2 h^{-2}- \lambda^{-2} & h^{-2} & 0 & .\\ h^{-2} & -2 h^{-2}- \lambda^{-2} & h^{-2} & 0 & .\\ . & . & . & .&. \\. & . & . & . & .\end{pmatrix}
 =\begin{pmatrix} 
 v & -w & 0 & .\\  0 & v & -w & 0\\ . & . & . & . \\. & . & 0 & v \end{pmatrix} \cdot
 \begin{pmatrix} v & 0 & . & .\\  -w & v & 0 & .\\ . & . & . & . \\. & 0 & -w & v \end{pmatrix}
 \nonumber \\
&=& {\bf D} \cdot {\bf D}^{\rm T} = \begin{pmatrix}  v^{2}+w^{2} & -vw & 0 & . &.\\  -vw & v^{2}+w^{2} & -vw & 0 & .\\ 0 & -vw & v^{2}+w^{2} & -vw & 0 \\ . & . & . & .\\.& . & 0 & -vw & v^{2} 
 \end{pmatrix},
\end{eqnarray}
which gives the following pair of equations for the elements of the square-root matrix, $v$ and $w$:
$v^{2} + w^{2} = 2 h^{-2} + \lambda^{-2}$, \& $v \cdot w = h^{-2}$. The roots simplify in the continuum limit to,
\begin{equation}
v,w = \frac{1}{2} \big( \sqrt{\lambda^{-2} + 4 a^{-2}} \pm \lambda^{-1} \big) \; or \; v \simeq \frac{1}{h} + \frac{1}{2 \lambda} + \mathcal{O}(h^{2}) \, , w \simeq \,  \frac{1}{h} + \mathcal{O}(h^{2})
\end{equation}
for the limit, $\lambda \gg h$ in which the continuum theory is valid.

There is clearly a degeneracy in the roots  $v$ \& $w$ above (the values of $v$ and $w$ can be switched), but we now show that the boundary condition specifies a particular choice that corresponds to the existence of the  zero mode localized at this boundary.  In order to obtain a zero mode for the matrix ${\bf D}^{\rm T}$, such that ${\bf D}^{\rm T} |u \rangle = 0$, we consider the general eigenvalue problem: $ {\bf D}^{\rm T} | \epsilon \rangle = \epsilon | \epsilon \rangle $:
\begin{equation}
\begin{pmatrix} v & 0 & . & .\\  -w & v & 0 & .\\ . & . & . & . \\. & 0 & -w & v \end{pmatrix}
\begin{pmatrix} u_{1} \\ . \\. \\ u_{N} \end{pmatrix} = \epsilon \begin{pmatrix} u_{1} \\ . \\. \\ u_{N}  \end{pmatrix}.
\end{equation}
The $N^{th}$ row of the matrix ${\bf D}^{\rm T}$ enforces a  boundary condition on its zero eigenvector:  $v u_{N} - w u_{N-1} =0$. If we expect to find a zero mode exponentially localized at the $N_{th}$ site, the condition: $u_{N-1} < u_{N}$ has to be satisfied, \emph{i.e.}  $v<w$. This picks out a particular choice for the values of $v$ and $w$, which yields a localization length, $h \log(w/v) \simeq \lambda$, consistent with the solution of the hydrodynamic equation.
%Show numerically that a particular choice of $v$ and $w$ leads to z zero eigenvector with the expected localization length, $\lambda$. 

We now show that the bulk of the matrix, ${\bf S}$, can be characterized by a winding number which is related to the presence of a localized zero mode at the boundary through an index theorem \cite{Kane2013}.  With periodic boundary conditions, the bulk can be represented in Fourier basis as $D(k) = v - w e^{i k}$. The winding number, $ \nu = 1/(2 \pi) \int dk \frac{d}{dk} \arg(D(k))$, is $\nu = 1$ when $v < w$ and $\nu = 0$ when $v > w$.  This suggests that the presence of a boundary mode requires  $v<w$ as we showed explicitly.  A crossover is induced from the topological to the trivial phase when $v=w$, expected for the case $\lambda^{-1} = 0$ \emph{i.e.} the decay length scale diverges. This happens in the hydrodynamic theory when there is zero substrate friction and corresponds to the closing of the gap in the eigenvalue spectrum (as shown in Fig.3d in the main text).

Finally, following Ref.~\cite{Kane2013}, we can obtain intuition for how the presence of friction leads to an effective polarization by constructing a spring like analogue that has a eigenvalue structure similar to the dissipative hydrodynamic operator. Such a description is detailed in SI Fig.~\ref{fig:spring} and consists a linear elastic chain attached to a substrate with additional springs. This substrate-string interaction models the role of friction. 
As detailed in the figure, this elastic system supports no edge modes when the vertical springs connecting the elastic chain to the substrate exert no force on average. In cases where the vertical springs exert a force on average \textendash this mimics the role played by friction \textendash the system supports robust localized edge modes. The asymmetry between the coefficients of the SSH model for this spring system is proportional to $\cot(\theta)$. The construction in SI Fig.~\ref{fig:spring} helps clarify the origin of polarization in out of equilibrium dissipative systems by constructing a topologically nontrivial mechanical system.

\subsection*{Supp. Note 4: Steady state solution of rotor hydrodynamics}
We discuss here the solution of the hydrodynamic equations for actively spinning rotors, Eq.(15), in the main text.
The steady state hydrodynamic equations for spin and vorticity of driven rotors are given by \cite{vanZuiden16, Tsai2005},
\begin{eqnarray}
(\nabla^{2} - \lambda_{\Omega}^{-2}) \Omega + a \omega =  -\tau/D_{\Omega}, \\ \nonumber
(\nabla^2 -\lambda_{\omega}^{-2})\omega - b \nabla^{2} \Omega = 0,
\label{rotordyn_ss}
\end{eqnarray}
where, $\lambda_{\Omega}^{-2} = (\Gamma + \Gamma_{\Omega})/D_{\Omega}$,
and, $ \lambda_{\omega}^{-2} =  \Gamma_{v}/(\eta + \Gamma)$,
$a = \Gamma/D_{\Omega}$, $b = \Gamma/(\eta + \Gamma)$, are quantities that depend on material parameters in the hydrodynamic theory and are defined after Eq. 16 in the main text. The above equations have a spatially homogeneous steady state which describes the bulk: $ \Omega_{\rm{b}} = \tau/(\Gamma+ \Gamma_{\Omega})$ and zero vorticity.  
We look for spatially localized steady states.  To do this, we redefine:  $\Omega$ as $\Omega - \Omega_{\rm{b}}$,
and integrate out vorticity, $\omega$, to obtain the 4th order differential equation,
\begin{equation}
\nabla^{4} \Omega - c \lambda_{\Omega}^{-2} \nabla^{2} \Omega
 +  \lambda_{\Omega}^{-4} s^{2} \Omega = 0,
 \label{sseq_Omega}
 \end{equation} 
% $ c =(\lambda_{\Omega}^{-2} + \lambda_{\omega}^{-2} - a \cdot b)$
where, $s = \lambda_{\Omega}/\lambda_{\omega}$, $c= 1 + s^{2} - ab \lambda^{2}_{\Omega}$. 
The roots of the quartic equation resulting from $\Omega(x) = \exp(-t x)$ 
give the two decay length scales,
\begin{equation}
2\lambda^{-2}_{S,L} = \lambda_{\Omega}^{-2} (c \pm \sqrt{c^{2}-4 s^{2}}),
\label{quart_sol}
\end{equation}
There are {\emph two} independent modes, $\exp(\pm x/\lambda_{S})$ \& $ \exp(\pm x/\lambda_{L})$: the general solution is given by their linear superposition with coefficients that are determined by boundary conditions of the physical problem.  

%Since $b = \Gamma/(\eta + \Gamma) < 1$, $a \lambda^{2}_{\Omega} = \Gamma/(\Gamma + \Gamma_{\Omega}) \le 1$, $c > s$.   real roots to Eq. \ref{quart_sol} exist when $c> 2 s$, or $ 1- s>   \Gamma/\sqrt{(\Gamma + \Gamma_{\Omega})(\eta + \Gamma)} $.  As long as substrate friction is small, such that $\lambda_{\omega}$ remains significantly greater than $\lambda_{\Omega}$, there is a large range of values of spin-vorticity coupling strength for which Eq.\ref{quart_sol} has two real solutions. 
 
 We follow Ref.~\cite{vanZuiden16} in considering the low spin-vorticity and low substrate friction limit, $\Gamma, \Gamma_{v} \ll \eta$, which leads to the simplification:  $s = \lambda_{\Omega}/\lambda_{\omega} \ll 1$, $c \simeq 1+s^{2}$, and therefore, the decay lengths in Eq.\ref{quart_sol} decouple to $\lambda_{\omega}$ and $\lambda_{\Omega}$.  In this low $s$ limit, the spin reaches its bulk value as suggested by Eq.~\ref{sseq_Omega} away from the boundaries, whereas it is constrained to be zero at the boundary.  Therefore in a slab geometry that extends between: $-L/2 < x < L/2$, the spin velocity profile that is consistent with these boundary conditions has the approximate form:
\begin{equation}
\Omega(x) \simeq \Omega_{\rm{b}} \bigg[1 - \sech\bigg( \frac{L}{2 \lambda_{\Omega}} \bigg) \cosh \bigg( \frac{x}{ \lambda_{\Omega}} \bigg) \bigg]
\end{equation},
where $\Omega_{\rm{b}} = \tau/(\Gamma + \Gamma_{\Omega})$ is the bulk value of spin that a dilute gas of driven rotors would have, and where we have used the approximation that the spatial variation is dominated by the smaller decay length, $\lambda_{\Omega}$.

Since the rotors retract from the confining boundaries in the molecular simulations (see Fig. 4 in the main text), a likely boundary condition is zero tangential force at the wall: $\partial_{x} v_{y} + \Gamma (\Omega - \omega) \vert_{x=\pm L/2} = 0$, which translates to zero vorticity at the walls. Knowing the spin profile, the vorticity profile is determined by the linear momentum balance in Eq. S14b :
\begin{equation}
\omega(x) \simeq  \frac {b \Omega_{\rm{b}}}{1-\lambda_{\Omega}^{2}/\lambda_{\omega}^{2}} \bigg[ \sech\bigg( \frac{L}{2 \lambda_{\omega}} \bigg) \cosh \bigg( \frac{x}{ \lambda_{\omega}} \bigg) - \sech\bigg( \frac{L}{2 \lambda_{\Omega}} \bigg) \cosh \bigg( \frac{x}{ \lambda_{\Omega}} \bigg)  \bigg]
\end{equation}
and the velocity is given by,
\begin{equation}
v_{y} (x) =  \frac {b \Omega_{\rm{b}}}{(1-\lambda_{\Omega}^{2}/\lambda_{\omega}^{2})}
 \bigg[ \lambda_{\omega}  \sech\bigg( \frac{L}{2 \lambda_{\omega}} \bigg) \sinh \bigg( \frac{x}{ \lambda_{\omega}} \bigg) -  \lambda_{\Omega} \sech\bigg( \frac{L}{2 \lambda_{\Omega}} \bigg) \cosh \bigg( \frac{x}{ \lambda_{\Omega}} \bigg)  \bigg].
\label{vnostress}
\end{equation}
Additionally, one can propose a no-slip boundary condition at the wall, which results in the velocity profile:
\begin{equation}
v_{y} (x) =  \frac {b \Omega_{\rm{b}} \lambda_{\omega}}{(1-\lambda_{\Omega}^{2}/\lambda_{\omega}^{2})} \sinh \bigg(\frac{L}{2 \lambda_{\omega}}\bigg)
\bigg[  \csch\bigg( \frac{L}{2 \lambda_{\omega}} \bigg) \sinh \bigg( \frac{x}{ \lambda_{\omega}} \bigg) - \csch\bigg( \frac{L}{2 \lambda_{\Omega}} \bigg) \cosh \bigg( \frac{x}{ \lambda_{\Omega}} \bigg)  \bigg].
\label{vnoslip}
\end{equation}

\subsection*{Supp. Note 5: Bulk properties of rotor hydrodynamical operator}
The time-evolution of spin and vorticity, $ d/dt \ket{\Omega,\omega} = {\bf M} \ket{\Omega,\omega}$, fields is controlled by the dynamical operator, ${\bf M}$, (Eq.~12 in the main text), which is represented in a Fourier basis:

\begin{equation}
\frac{d}{dt}  \begin{pmatrix} \Omega \\ \omega \end{pmatrix} = -
\begin{pmatrix} 
 k^2 + \lambda_{\Omega}^{-2} & -a\\
 -b k^{2} &  k^2 +\lambda_{\omega}^{-2}
\end{pmatrix}
\begin{pmatrix} \Omega \\ \omega \end{pmatrix}, 
\end{equation}
where $a$ and $b$ contain the parameters of the hydrodynamic theory as defined in Eq.~(\ref{rotordyn_ss}).
The $2 \times 2$ dynamical evolution operator defined above , ${\bf M(}k)$, has positive eigenvalues, and can be symmetrized by a similarity transformation, $ M_{\rm{R}}(k) = {\bf T}^{-1}(k) {\bf M}(k) {\bf T}(k)$,
where the transformation matrix is, ${\bf T}(k) = \sqrt{k} \begin{pmatrix} 
(a/b)^{1/4} & 0\\
0 &  (b/a)^{1/4}
\end{pmatrix}$.
Note that an upper cutoff length scale (corresponding to system size) ensures that the inverse of ${\bf U}$ is well-defined.  This ``rotated'', symmetrized operator (with the same set of eigenvalues) is given by, ${\bf M_{\rm{R}}}(k) = -\begin{pmatrix} 
 k^2 + \lambda_{\Omega}^{-2} & k \sqrt{ab}\\
k \sqrt{ab} &  k^2 + \lambda_{\omega}^{-2}
\label{mat_rot}
\end{pmatrix}$.
This matrix can be factorized as ${\bf M_{\rm{R}}} = {\bf D} {\bf D}^{\rm T}$ using the properties of symmetric positive semi-definite matrices.  This follows from a construction based on symmetric matrices being diagonalized by unitary operators: ${\bf M_{\rm{R}}} = {\bf U}^{-1} {\bf M_{D}} {\bf U} = ({\bf U}^{-1} 
 \cdot \sqrt{{\bf M_{D}}} \cdot{\bf U}) \cdot({\bf U}^{-1}\cdot \sqrt{{\bf M_{D}}}\cdot{\bf
 U}) = {\bf D} {\bf D}^{T}$, where ${\bf D} = ({\bf U}^{-1}\cdot\sqrt{{\bf M_{D}}}\cdot{\bf U})$, and $\sqrt{{\bf M_{D}}}$ is a diagonal matrix whose elements are the square-root of eigenvalues of ${\bf M_{\rm{R}}}$. ${\bf U}$ is the unitary operator that diagonalizes ${\bf M_{\rm{R}}}$ to its diagonalized form,${\bf M_{D}}$ . 

%This factorization can be carried out in any basis, both in real and Fourier space and establishes that the hydrodynamic operator can be transformed to have the form, ${\bf D} {\bf D}^{T}$.

This provides a route to construct the square-root matrices, ${\bf D}$ and ${\bf D}^{\rm T}$ given a symmetric positive semi-definite matrix, in this case, ${\bf M_{\rm{R}}}$. The explicit form of ${\bf D}$ for the ${\bf M_{\rm{R}}}$ corresponding to  rotor hydrodynamics in the bulk can be calculated after some tedious algebra but provides little insight as to its topological properties or analogies with simple models of topological insulators, such as the SSH model. However, this construction proves that the ``rotated'' hydrodynamic operator for the active spinners can indeed be factorized into the form, ${\bf D} {\bf D}^{\rm T}$, and can therefore be mapped to a Hermitian operator for which a topological index theorem exists.

\subsection{Supp. Note 6: Bulk-boundary correspondence for the steady state of rotor hydrodynamics}

We discuss a route to establishing the notion of topological protection for the general spin-vortex coupled hydrodynamic equations in the previous section.  Here, we focus on the steady state of the hydrodynamic equations. This provides a more illuminating route to a topological index, besides allowing a clearer mapping to a general SSH-like tight-binding model.  We therefore factorize the steady state operator (obtained after integrating out $\omega$) corresponding to the $4^{th}$ order differential equation in Eq.~(\ref{sseq_Omega}).
This allows us to define a matrix ${\bf S_{s}}$ (which is diagonalized in Fourier basis with periodic boundary conditions) as,
\begin{equation}
{\bf S_{s}} =
\begin{pmatrix} 
0 & {\bf D}_{\rm s}\\
{\bf D}_{\rm s}^{\rm T} & 0
\end{pmatrix}
= \begin{pmatrix} 
0 & v+ w_{1} e^{i k} + w_{2} e^{ 2 i k} \\  v + w_{1} e^{-i k} + w_{2} e^{- 2 i k} & 0 
\end{pmatrix} 
\end{equation}
This corresponds to a SSH model with next nearest neighbor hopping (from one sublattice to another in the neighboring unit cell) with rate  $w_{2}$. Its winding number is computed according to Eq.~17 of the main text, and depends on the relative values of $v$, $w_{1}$ \& $w_{2}$:
\begin{equation}
\nu = \frac{1}{2\pi i} \int_{0}^{2 \pi}  dk \, \frac{d}{dk} \log \, \det S(k) =
\begin{cases}
0 &\text{for $v \geq w_{1} + w_{2}$},
\\
1 &\text{for $v < w_{1} - w_{2}$},
\\
2 &\text{for $|w_{2} - w_{1}| < v < w_{1} + w_{2} $},
\end{cases}
\end{equation}
where the matrix elements, $v$, $w_{1}$ and $w_{2}$ depend on the parameters of the differential operator in Eq.~(\ref{sseq_Omega}):  $\lambda_{\Omega}$, $c$ and $s$. The expressions for $v$, $w_{1}$ and $w_{2}$ are obtained after factorizing the discretized real space representation of the steady state differential operator, ${\bf L}_{\rm{s}} $ as discussed below.

Although the bulk properties of the hydrodynamic operator are easily seen in the Fourier basis, a demonstration of bulk-edge correspondence requires a discretized real space representation. The steady state hydrodynamic operator for $\Omega$ is discretized as a $N \times N$ matrix, ${\bf L}_{\rm s}$.  This $4^{th}$ order differential operator when discretized in real space, is represented by a sparse matrix which is nonzero only in the diagonal, and its two nearest bands on either side of the diagonal. These elements are labeled, $d$, $n_{1}$ and $n_{2}$ below.  We now show explicitly the decomposition of this dynamical operator matrix into ${\bf L}_{\rm s}= {\bf D}_{\rm s} {\bf D}_{\rm s}^{T}$ form, where $(D_{s})_{ij} = v \delta_{i,j} - w_{1} \delta_{i-1,j} + w_{2} \delta_{i-1,j}$ is a sparse upper triangular matrix with only the diagonal and its two nearest upper bands:  
\begin{equation}
{\bf L}_{\rm s} =\begin{pmatrix} 
  d & -n_{1} & n_{2} & 0 & . &.\\ -n_{1} & d & -n_{1}& n_{2} & 0 & .\\  n_{2} & -n_{1} & d & -n_{1} & n_{2} & 0\\ . & . & . & . &. &.  \\.&. & . & . & . & . \end{pmatrix} = \begin{pmatrix} 
   v & -w_{1} & w_{2} & 0 & .\\  0 & v & -w_{1} & w_{2} & 0\\ 0& 0 & v & -w_{1} & w_{2} \\ . & . & . & . & . \\.& . & . & 0 & v \end{pmatrix} \cdot
   \begin{pmatrix} v & 0 & . & . & .\\  -w_{1} & v & 0 & . & .\\ w_{2} & -w_{1} &  v & 0 & .\\ . & . & . & .  & .\\. & 0 & w_{2} & -w_{1} & v \end{pmatrix}.
   \nonumber \\
\end{equation}

The terms of the factorized matrix, ${\bf D}$, are given by:
\begin{eqnarray}
n_{2} &=& v \cdot w_{2} = h^{-4}, \nonumber \\ 
n_{1} &=& w_{1} \cdot (v + w_{2}) = 4 h^{-4} + c h^{-2} \lambda^{-2}_{\Omega},  \nonumber \\ 
d &=& v^{2} + w_{1}^{2} + w_{2}^{2} = 6  h^{-4} + 2 c h^{-2} \lambda^{-2}_{\Omega} + s^{2} \lambda^{-4}_{\Omega}.
\label{eq_vw1w2}
\end{eqnarray}

These algebraic equations for $v$, $w_{1}$ and $w_{2}$ in terms of the grid mesh size, $h$, and the parameters, $c$, $s$ and $\lambda_{\Omega}$ of the hydrodynamic theory can be solved analytically. Note that $v$ \& $w_{2}$, and $(v + w_{2})$ \& $w_{1}$ can be interchanged in Eq.~(\ref{eq_vw1w2}) without changing the equations. 
% This points to a degeneracy in the choices of $v$, $w_{1}$ and $w_{2}$. 
We show below that the right choice that is consistent with two localized zero eigenvectors at the boundary can be written in approximate form for $h \ll \lambda_{\Omega}$ (the mesh size is much smaller than the decay length in a hydrodynamic theory) as,
\begin{eqnarray}
w_{1} &\simeq& 2 h^{-2} \Big[ 1 + \frac{1}{8} (c - 2 s) \bigg( \frac{h}{\lambda_{\Omega}} \bigg)^{2}  \Big], \nonumber \\
v &\simeq& h^{-2} \Big[1 - \frac{1}{2} \sqrt{c + 2 s}   \frac{h} {\lambda_{\Omega}} + \frac{1}{8} (c + 2 s)\bigg( \frac{h}{\lambda_{\Omega}} \bigg)^{2}  \Big], \nonumber \\
w_{2}  &\simeq&  h^{-2} \Big[1 + \frac{1}{2} \sqrt{c + 2 s}  \frac{h} {\lambda_{\Omega}} + \frac{1}{8} (c + 2 s)\bigg( \frac{h}{\lambda_{\Omega}} \bigg)^{2} \Big].
\label{sol_vw1w2}
\end{eqnarray}
Now consider a zero mode of ${\bf D}_{\rm s}^{\rm T} $:
\begin{equation}
\begin{pmatrix} v & 0 & . & . & .\\  -w_{1} & v & 0 & . & .\\ w_{2} & -w_{1} &  v & 0 & .\\ . & . & . & .  & .\\. & 0 & w_{2} & -w_{1} & v \end{pmatrix} 
\begin{pmatrix} u_{1} \\ . \\. \\.\\ u_{N} \end{pmatrix} = \epsilon \begin{pmatrix} u_{1} \\ . \\. \\. \\ u_{N} \end{pmatrix}. 
 \end{equation}
In order for the zero eigenfunction to be exponentially localized at the boundary corresponding to the $N_{th}$ site, that is, for $u_{N-1}/u_{N} = u_{N-2}/u_{N} = r$ (where $r = \exp (-h/\lambda) <1$) to be satisfied, we require:
\begin{equation}
r^2- (w_{1}/w_{2}) r + v/w_{2} =0
\end{equation}
The two roots of this equation in $r$ correspond to the two decay length scales characterizing the two localized modes that the solution of the hydrodynamic equations predict.  By inserting the values for $v$, $w_{1}$ and $w_{2}$ displayed in Eq.~(\ref{sol_vw1w2}) into the solution of the quadratic equation, we recover $\lambda_{1,2} = -h/\log(r)  \simeq 2 \lambda_{\Omega}/ (\sqrt{c+2s} \pm \sqrt{c-2s})$ which are the same two decay length scales obtained from the solution of the steady state rotor hydrodynamics in Eq. 16 \cite{Tsai2005}.  This indicates that this is the right choice of solution for $v$, $w_1$ and $w_2$ that satisfies the matrix decomposition ${\bf L}_{\rm s}= {\bf D}_{\rm s} {\bf D}_{\rm s}^{\rm T}$ in the bulk and also supports the two localized boundary modes we know ought to exist from the physics of the hydrodynamic problem.  We see from Eq.~(\ref{sol_vw1w2}) that these values satisfy $w_{2} < v < w_{1}$ which corresponds to a winding number of $\nu = 2$.  We have thus shown the topological bulk-boundary correspondence for these equations.

\bibliography{References}